\documentclass[aps,showkeys,physrev,reprint,superscriptaddress,twocolumn]{revtex4-2}

\usepackage{graphicx}
\usepackage{bm}
\usepackage{xcolor}
\usepackage{amsmath}
\usepackage{makecell}

\usepackage[normalem]{ulem}
\newcommand{\nt}{\notag}
\newcommand{\del}{\partial}

\makeatletter
\renewcommand*{\@fnsymbol}[1]{\ensuremath{\ifcase#1%
  \or ~
  \or \dagger
  \or \ddagger
  \or \star
  \or \nabla
  \else\@ctrerr\fi}}
\makeatother

\begin{document}
\title{Visible and Terahertz Nonlinear Responses in the Topological Noble Metal Dichalcogenide PdTe$_2$}

\email[$\dagger$ $\ddagger$ ]{These authors contributed equally to this work.}

\author{George J. de Coster}
\email[]{george.j.decoster.civ@army.mil}
\affiliation{Institute of Physics and Center for Integrated Sensor Systems (SENS), University of the Bundeswehr Munich, Werner-Heisenberg-Weg. 39, 85577 Neubiberg, Germany}
\affiliation{DEVCOM Army Research Laboratory, 2800 Powder Mill Road, Adelphi, MD, 20738 USA}

\author{Lucas Lafeta}
\email[]{lucas.lafeta@cup.lmu.de}
\affiliation{Department of Chemistry and Center for NanoScience (CeNS), Ludwig- Maximilians-Universität München, Butenandtstraße 5-13 (E), 81377 Munich, Germany}

\author{Stefan Heiserer}
\affiliation{Institute of Physics and Center for Integrated Sensor Systems (SENS), University of the Bundeswehr Munich, Werner-Heisenberg-Weg. 39, 85577 Neubiberg, Germany}

\author{Cormac Ó Coileáin}
\affiliation{Institute of Physics and Center for Integrated Sensor Systems (SENS), University of the Bundeswehr Munich, Werner-Heisenberg-Weg. 39, 85577 Neubiberg, Germany}

\author{Zdenek Sofer}
\affiliation{Department of Inorganic Chemistry, University of Chemistry and Technology Prague, Technická 5, 166 28 Prague 6, Czech Republic}

\author{Achim Hartschuh}
\affiliation{Department of Chemistry and Center for NanoScience (CeNS), Ludwig- Maximilians-Universität München, Butenandtstraße 5-13 (E), 81377 Munich, Germany}

\author{Georg S. Duesberg}
\affiliation{Institute of Physics and Center for Integrated Sensor Systems (SENS), University of the Bundeswehr Munich,  Werner-Heisenberg-Weg. 39, 85577 Neubiberg, Germany}

\author{Paul Seifert}
\email[]{paul.seifert@unibw.de}
\affiliation{Institute of Physics and Center for Integrated Sensor Systems (SENS), University of the Bundeswehr Munich,  Werner-Heisenberg-Weg. 39, 85577 Neubiberg, Germany}

\date{\today}

\begin{abstract}
Nonlinear processes can offer pathways to next-generation sensors and frequency mixing devices to overcome modern imaging, detection, and communication challenges. In this article, we report on strong second and third-order nonlinear optical responses in visible and terahertz (THz) light in single crystals of the noble metal dichalcogenide PdTe$_2$. We find that buried conduction and valence topological surface states of PdTe$_2$ lead to resonant optical second-harmonic generation.
%, similar to what has previously been observed in topological insulators. 
On the other hand, although the nonlinear responses obtained with THz excitation are not close to this resonance, they can be clearly observed in reflection geometry, even in the presence of broadband excitation, where optical filters are not necessary to observe the \textcolor{black}{enhanced} odd-order higher harmonic \textcolor{black}{output}. By carefully considering the radiative photocurrent framework of stimulated THz emission, we are able to extract fingerprints of both second- and third-order processes in the THz regime, and show that PdTe$_2$ is a promising material candidate for radio frequency rectification, frequency mixing, and beam focusing.
\end{abstract}
%\keywords{nonlinear optics, Palladium ditelluride, noble metal dichalcogenides, SHG, FWM}
\maketitle

\section{Introduction}
Quantum materials with strong spin–orbit coupling and topological surface states have spawned new device concepts that exploit nonlinear optical (NLO) rectification and higher-harmonic generation \cite{annrev2021,Shi2023,Sodemann2015,Onishi2024}, the Josephson diode effect \cite{Wu2022,Sivakumar2024}, and spin–orbit torque switching \cite{Liu2020,Hidding2023}. NLO responses in quantum materials are particularly relevant to polarization sensitive imaging, frequency conversion, and radio frequency (RF) detection, as even inversion symmetric crystal systems can exhibit strong second-harmonic generation (SHG), polarization dependent photocurrent generation, and rectification-processes classically forbidden by symmetry \cite{annrev2021,Connelly2024,Plank2018,Xie2025,Hemmat2023}. Such effects often originate from topological surface states, which enable broadband photocurrent and DC rectification spanning near-UV to THz frequencies \cite{Pan2017,Plank2018,Hu2025,Zhang2021}.

Layered noble metal dichalcogenides (NMDs), (Pd, Pt)(Te,Se,S)$_2$, have recently been recognized as a distinctive class of 2D quantum materials, which combine metallic conductivity, strong spin–orbit coupling, and topologically nontrivial electronic structures. They crystallize predominantly in the centrosymmetric 1T phase (space group P$\bar{3}$m1) with octahedral coordination of the transition metal (Fig.~1(a) and (b)), and have recently been shown to be accessible via low-temperature synthesis methods \cite{Yim2016,McManus2020}. Among them, PdTe$_2$ is a type-II Dirac semimetal which hosts both topological surface states and superconductivity \cite{zheng_2018,clark_2018}. Related compounds such as PtTe$_2$ and PtSe$_2$ exhibit non-saturating magnetoresistance, high carrier mobilities, and thickness-driven band-structure evolution from semimetallic to semiconducting behavior \cite{AliSciRep2018,SereniSciRep2016,Zhussupbekov2021,BalicasArXiv2018,PtSe2PRL2020,KrasheninnikovACSNano2021,PtTe2Crossover2019}.
Despite their inversion symmetry, NMDs display unexpectedly strong NLO effects, including SHG, shift currents, and photogalvanic responses, attributed to local symmetry breaking, surface and interface fields, or strain-induced distortions \cite{Faizanuddin2024,Wang2019,Wang2015,Chu_2024,Guo_2020}. Their large second-order susceptibilities are enhanced by spin–orbit coupling and Berry curvature effects, making them promising candidates for tunable NLO devices. Their 2D nature allows straightforward integration into MEMS architectures and tuning via thickness control, Janus engineering, or strain \cite{Shi2023,Yim2018,Peng2020,Ge2023,Cai2025,Heiserer2025}.

In this work, we verified the crystalline quality and symmetry of the PdTe$_2$ flakes using Raman spectroscopy (Fig.~1(c) and (d)), showing the characteristic $E_g$ and $A_{1g}$ phonon modes, and by lattice-resolution atomic force microscopy (AFM) imaging (Fig.~1(e)), revealing the expected hexagonal surface lattice (0.39~nm lattice constant). A representative bulk crystal is shown in Fig.~1(c), and the schematic band structure (Fig.~1(f)) illustrates the type-II Dirac dispersion with topological surface states and surface-projected Dirac cones.

\begin{figure*}[t!]
        \begin{center}
             \includegraphics[width=2\columnwidth]{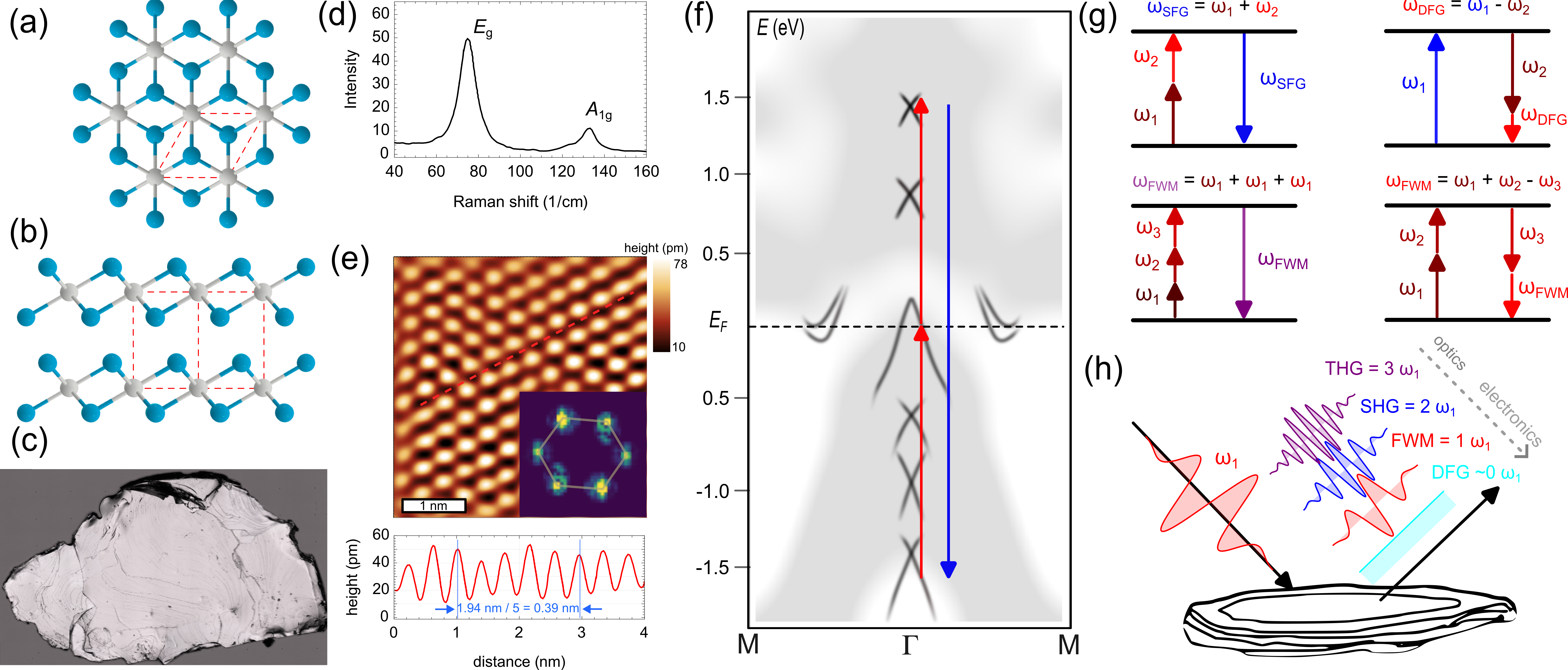}
        \end{center}
    \caption{(a)Top view and (b) side view of 1T layered crystal structure of PdTe$_2$ with Pd atoms shown in gray and Se atoms shown in blue. (c) Optical microscopy image of a $\sim$5 mm bulk layered PdTe$_2$ crystal. (d) Raman spectrum of PdTe$_2$. (e) Lattice-resolution AFM image (Fourier-filtered), with profile along the red dashed line (below). Inset, fast Fourier transform of the AFM image. (f) Schematic electronic structure of PdTe$_2$ along the $\Gamma-\text{M}$ direction with bulk bands indicated in gray and surface (projected) bands and topological surface states indicated in black. A possible resonant condition for SHG is shown by the arrows (red-excitation, blue-emission). (g) Generic second- and third-order sum-frequency generation (SFG), second-order difference-frequency generation (DFG), and third-order FWM processes, highlighting the energy exchange between interacting photons.(h) Schematic illustration of SHG, THG, FWM and DFG processes from a near monochromatic wave packet.
}
\label{fig1} 
\end{figure*}

Moreover, we demonstrate NLO and harmonic generation in PdTe$_2$  in both the visible ($\sim$1.5~eV) and terahertz range ($\sim$0.1-3 THz), revealing persistent and tunable nonlinear responses in this centrosymmetric topological semimetal. These responses at disparate optical bands invite a closer examination of the nonlinear mechanisms that can arise in layered systems such as PdTe$_2$. We outline the fundamental principles of NLO interactions in 2D materials, emphasizing how local symmetry breaking and spin–orbit–driven band topology can enhance their nonlinear susceptibilities.

\begin{figure*}[t!]
\begin{center}
 \includegraphics[width=2\columnwidth]{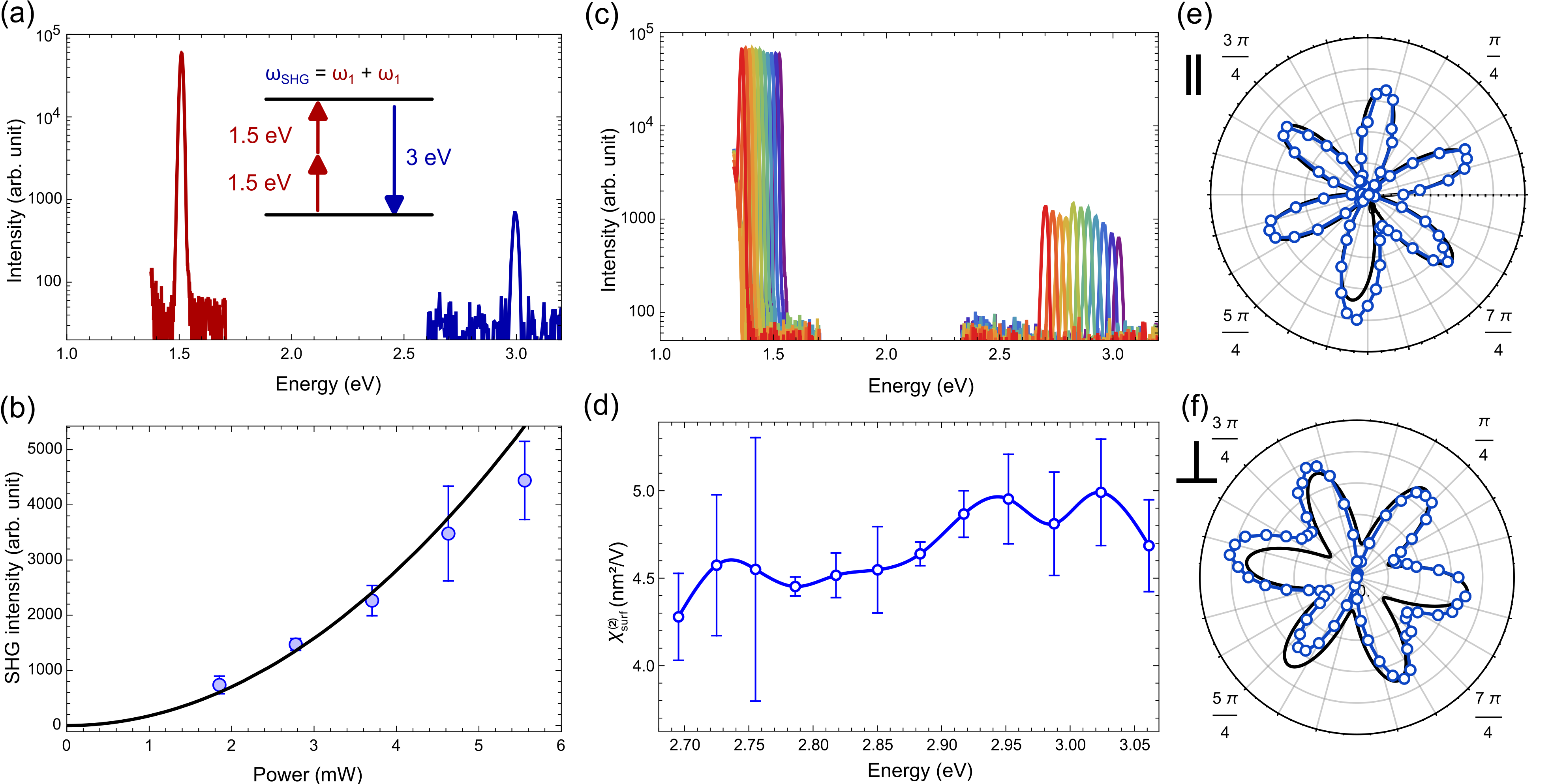}
\end{center}
 \caption{(a) Optical second harmonic generation at 3 eV in PdTe$_2$ stimulated by a 1.5 eV pulsed laser. (b) The power dependence of the SHG at 3 eV is shown to be quadratic in the pulse power, consistent with SHG scaling. (c) Spectra of excitation and SHG emission for varying excitation wavelength. (d) \textcolor{black}{Experimental values of PdTe$_2$ surface $\chi^{(2)}$, extracted by comparative method using crystalline quartz as a reference sample, for different wavelengths. } The symmetry of the parallel (e) and cross (f) polarization of the SHG response to varying the polarization of the 1.5\,eV pulse by a half-wave plate synchronous with a polarizer in front of the detector is checked and found to exhibit six-fold symmetry consistent with the C$_{3v}$ surface point group of PdTe$_2$.}
\label{fig2} 
\end{figure*}

\section{Principles of nonlinear optics in 2D materials}
NLO techniques have become essential tools for probing 2D materials. SHG, a second-order process, is widely used to determine lattice orientation and symmetry in noncentrosymmetric systems such as h-BN and transition metal dichalcogenides \cite{malard_observation_2013, li_probing_2013, lafeta_probing_2025}, while third-order techniques like four-wave mixing (FWM) probe resonant transitions and excitonic responses \cite{Lafeta_2021}.
The optical response of a medium arises from the induced polarization $\mathbf{P}(\mathbf{r},t)$ caused by the incident electric field $\mathbf{E}(\mathbf{r},t)$:
\begin{equation}
\mathbf{P} = \varepsilon_0 \left(\chi^{(1)} \mathbf{E} + \chi^{(2)} \mathbf{E}^2 + \chi^{(3)} \mathbf{E}^3 + \cdots \right),
\end{equation}
where $\chi^{(n)}$ is the $n$th-order susceptibility tensor. At low field intensities, the linear term ($\chi^{(1)}$) dominates, whereas higher intensities activate nonlinear terms. The tensor rank grows with order ($\chi^{(1)}$: 9 elements, $\chi^{(2)}$: 27, $\chi^{(3)}$: 81), and the nonvanishing elements are dictated by crystal symmetry \cite{Boyd}. Even-order processes vanish in perfectly centrosymmetric materials ($\chi^{(2n)}=0$), implying that SHG and related effects arise only when inversion symmetry is locally broken. In frequency space the second-order nonlinear polarization responsible for SHG is given by:
\begin{equation}
\mathbf{P}^{(2)}(2 \omega_1) = \varepsilon_0 \chi^{(2)} \mathbf{E}(\omega_1) \mathbf{E}(\omega_1) ~,
\end{equation}
which corresponds to the second-order sum frequency process in Fig. \ref{fig1}(g) with $\omega_1 = \omega_2$. The third-order term corresponding to FWM is:
\begin{equation}
\mathbf{P}^{(3)}(2 \omega_1 - \omega_3) = \varepsilon_0 \chi^{(3)} \mathbf{E}(\omega_1)\mathbf{E}(\omega_1)\mathbf{E}^*(\omega_3)~.
\end{equation}
This particular degenerate case of third-order corresponds to the bottom third-order optical process in Fig. \ref{fig1}(g) with $\omega_1 = \omega_2$. It gives rise to sum and difference degenerated FWM \cite{Boyd}. The spectral dependence of the nonlinear response tensors are discussed in Sections III and IV. We note here that enhancements to SHG and FWM are anticipated when the excitation pathways or emission correspond to resonances within the bandstructure \cite{Lafeta_2021}, as indicated by the resonant SHG process at 2.95 eV in Fig. 1(f). 

The form and symmetries of $\chi^{(2)}$ and $\chi^{(3)}$ are easily computed using the C$_{3v}$ surface point group of 1T-PdTe$_2$. These results are well known and not shown here \cite{Hsieh2011,McIver2012,ShenNLO}. When bulk inversion symmetry is further imposed on the tensors, $\chi^{(2)}$ becomes trivially zero.  In centrosymmetric materials like PdTe$_2$, strong spin–orbit coupling, surface states, and strain fields can locally break inversion symmetry and amplify Berry curvature effects, resulting in finite $\chi^{(2)}$ and enhanced $\chi^{(3)}$. These mechanisms provide a natural explanation for the broadband nonlinear and harmonic generation responses observed experimentally.

\section{Nonlinear Optical Responses to Visible Light in PdTe$_2$}
\subsection{Second Harmonic Generation}

The emission of SHG in noble metal dichalcogenides has been investigated previously \cite{Chu_2024, Yu_2021, Faizanuddin_arxiv_2024}. Despite the inversion symmetry of the point group $D_{3d}$ (P$\bar{3}$m1 space group) for the 1T phase (see Figs. 1(a) and (b)), experiments have seen non-zero emission of SHG by the surface of PdTe$_2$, governed by the C$_{3v}$ point group symmetries of the surface \cite{Faizanuddin_arxiv_2024}. In our studies, we focus the experiments on the PdTe2 sample shown in Fig.~\ref{fig1}(c). To probe second-harmonic generation, we excite PdTe$_2$ with a pulsed 1.5eV ($\sim$825nm) laser, generating a second-order signal at 3eV ($\sim$412.5nm). Figure \ref{fig2}(a) shows the observed spectra of the emission of the SHG in blue and the residual incident field after the short-pass in red.

The quadratic power dependence of the SHG intensity is shown in Fig.~\ref{fig2}(b) using the same condition as measured in Fig.~\ref{fig2}(a). This quadratic behavior is characteristic of the second-order process, thereby corroborating the second-harmonic generation in PdTe$_2$. Figure ~\ref{fig2}(c) presents the spectra of SHG using various wavelengths, the region at 1.5\,eV shows the incident field after short-pass filtering. The peaks near 3\,eV are SHG at the corresponding incident wavelengths.

\textcolor{black}{The second-order nonlinear susceptibility of PdTe$_2$ surface ($\chi^{(2)}$) was measured using a comparative method with crystalline quartz as a reference sample \cite{malard_observation_2013, Lafeta_2021} and the values of $\chi^{(2)}$ for different wavelengths are shown in Figure ~\ref{fig2}(d). The method used to obtain the $\chi^{(2)}$ value, and a comparison with other materials (such as monolayers of WS$_2$, MoS$_2$, and MoSe$_2$), are discussed in Appendix \ref{Appchi}} 

Figures ~\ref{fig2}(e) and (f) respectively show the experimental polar dependency of the SHG signal upon rotating the incident electric field synchronously with the polarizer in the detection parallel and perpendicular, respectively. The six-fold rotational symmetry of the polarization intensity is the signature of an underlying three-fold rotation symmetric lattice. The data is closely fit by the functional form (black) expected for surface  C$_{3v}$ point group symmetry \cite{Hsieh2011,McIver2012}. The defects, on the other hand, possess a generic two-fold rotational symmetry of their SHG spectrum (see Appendix \ref{AppA}) which follows from many point groups and thus only serves to show the defects break the symmetry of the parent lattice. 

 \begin{figure}[t!]
\begin{center}
 \includegraphics[width=1\columnwidth]{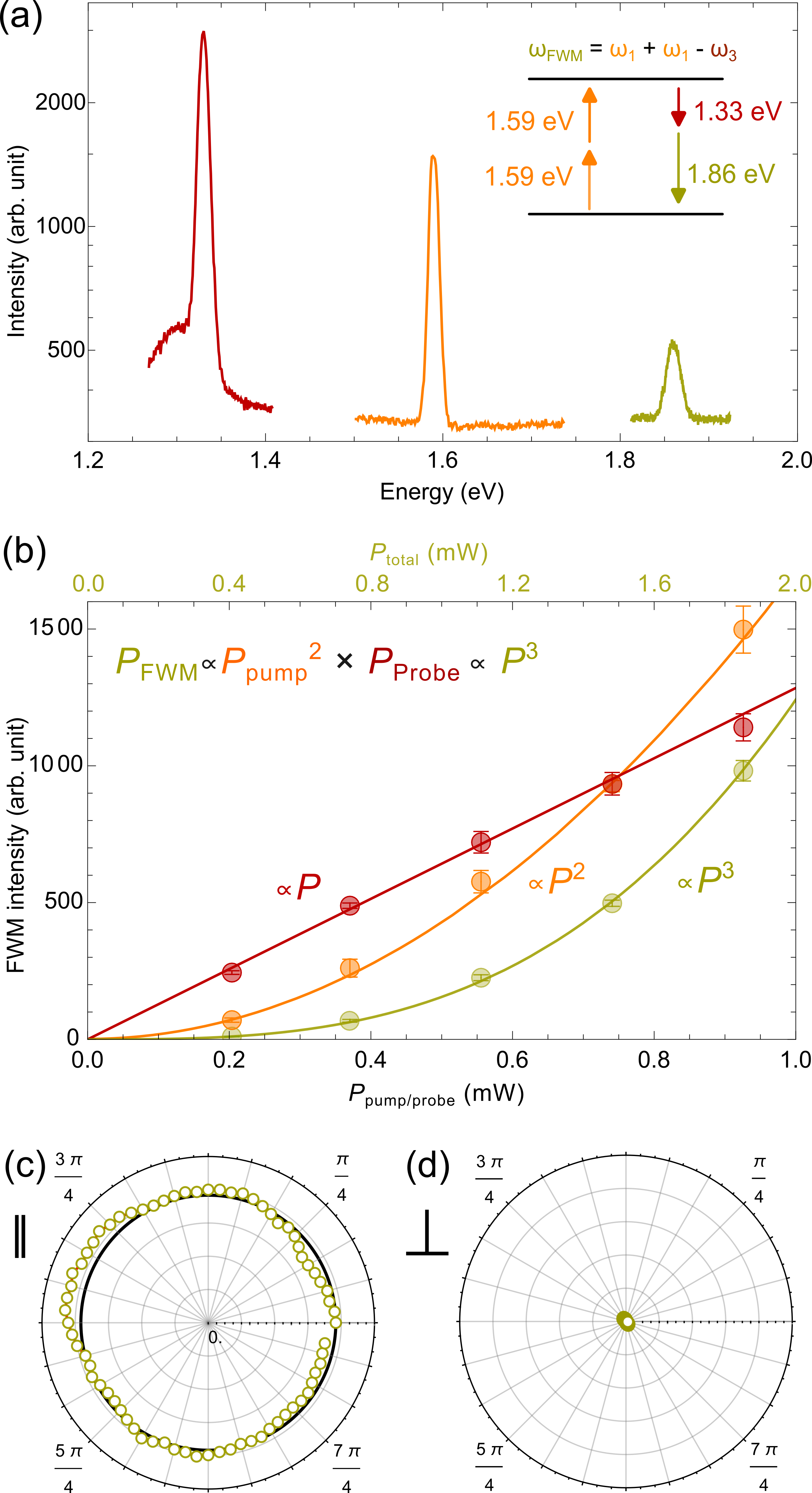}
\end{center}
 \caption{(a) Optical FWM probing of the third-order nonlinear response of PdTe$_2$ is conducted using 1.59eV and 1.33 eV pump and probe pulses. (b) The power dependence of the FWM intensity is quadratic in pump intensity and linear in probe intensity, yielding an overall cubic intensity dependence, consistent with a third-order effect. The input fields were subjected to a half-wave plate rotation of their polarization, and the symmetry of the parallel (c) and cross (d) polarization-detected fields was probed. Consistent with C$_{3v}$ surface point group symmetry, there is effectively no third-order response at cross polarization, and a constant signal for the parallel detection.} 
\label{fig3} 
\end{figure}

\begin{figure*}[t]
\begin{center}
 \includegraphics[width=2\columnwidth]{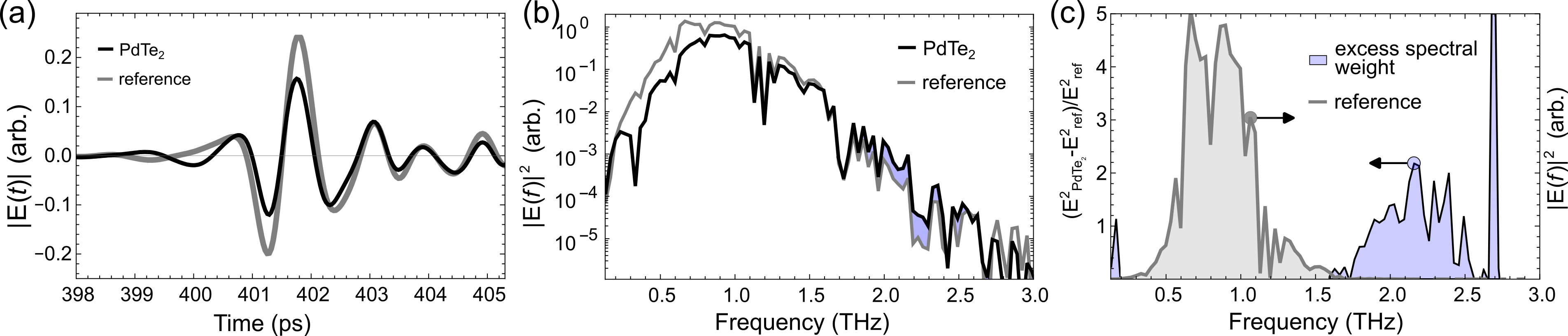}
\end{center}
 \caption{(a) Time scans of the reference injected THz pulse and that emitted by PdTe$_2$ after illumination. The pulses are overlaid to show their qualitative difference. Their intensity Fourier spectra in (b) show the PdTe$_2$ emission is greater than the reference for some higher and lower frequencies. (c) Excitation spectrum (gray) and \textcolor{black}{relatively enhanced spectral weight} in the reflected pulse (blue)}
\label{fig4} 
\end{figure*}

\subsection{Four-Wave Mixing}

We next investigated the third-order nonlinear regime using FWM, which, unlike second-order processes, does not rely on symmetry breaking and is allowed in all media. FWM provides direct access to the third-order susceptibility $\chi^{(3)}$, offering insight into the intrinsic electronic nonlinearity of PdTe$_2$. Special cases such as third-harmonic generation (THG) and degenerate FWM (DFWM) have been widely used to probe materials, including h-BN, graphene, carbon nanotubes, and TMDs \cite{Virga_2019, gordeev_2023, Lafeta_2017, Lange_2024, wen2019, autere_nonlinear_2018}.

We demonstrate a DFWM process in bulk PdTe$_2$ samples. Figure ~\ref{fig3}(a) shows the FWM spectrum generated by two degenerate pump fields at 1.59 eV ($\sim$780 nm) and one probe at 1.33 eV ($\sim$932 nm), producing a fourth wave at 1.86 eV ($\sim$667 nm). Fig.~\ref{fig3}(b) presents the power dependence of the FWM signal which follow the the proportion $I_{FWM}\propto I^2_{pump}I_{probe}$. 
When the pump power is fixed at $\sim1.1$ mW and the probe power is varied, the FWM intensity shows linear scaling with $I_{probe}$ (red curve). Inversely, by fixing the probe power at $\sim1.1$ mW and varying the pump power, the FWM intensity then scales as $I_{probe}^2$ (orange curve). The FWM intensity scaling can be calculated from the combination of these two measurements, and results in a cubic scaling with total input power (green curve).

We further probed the symmetries of the FWM signal by subjecting both pump and probe incident laser to half-wave plate rotations by synchronously rotating a polarizer in front of the detector, and then collecting the resulting signal at parallel and cross polarizations to the incident beams. The results in Figs. \ref{fig3}(c) and (d) show near-uniform behavior as a function of wave-plate angle, with almost no signal present in the cross-polarization detection. This is consistent with the symmetry of PdTe$_2$ for third-order processes. The slight deviations from uniformity in Fig. \ref{fig3}(c) can be attributed to the optical setup efficiency in certain polarization directions and alignment of the wave-plate with the polarizer in the detection.

The observation of a strong DFWM signal in PdTe$_2$ reveals a clear third-order nonlinear susceptibility ($\chi^{(3)}$). \textcolor{black}{The effective third-order nonlinear susceptibility can be extracted using a comparative method \cite{Lafeta_2021} (analogous to the second-order case shown earlier) with a well-established reference (in this case, fused quartz). The effective $\chi^{(3)}$ for a bulk of PdTe$_2$ found using the same wavelength as shown in Fig.~\ref{fig3}(a) was $\chi^{(3)}_{\mathrm{eff}}\approx2\cdot10^{-21}\,m^2/V_2^2$ (further information and discussion can be found in Appendix \ref{Appchi}).} This response can arise from contributions of Dirac-like interband transitions and surface electronic states, which potentially enhance $\chi^{(3)}$ even in metallic systems \cite{Mrudul2021, Mao2024, Mikhailov2014, Hafez2018, Tielrooij2022, Soavi2018}. This highlights PdTe$_2$ as a promising platform for broadband nonlinear photonics that extends beyond the family of semiconducting 2D materials.

\section{THz Spectroscopy}
By investigating the  response of PdTe$_2$ to THz radiation, we can determine if the NLO activity of interband transitions probed by SHG and FWM disappears for smaller, intraband energy scales (1\,THz corresponds to 4.14\,meV). Similarly to our experiments at optical frequencies, we performed THz time-domain spectroscopy on the PdTe$_2$ in atmosphere under reflection geometry using a pulse centered around $\sim$0.7\,THz. The reflected pulse was sampled in time, and a pulse reflected off a gold mirror was used as a reference signal to determine the spectrum of the excitation pulse.

The time-domain signal of a reference pulse and the THz pulse generated by reflecting off the PdTe$_2$ crystal surface are shown in Fig. \ref{fig4}(a). In Fig. \ref{fig4}(b) we show the spectrum of these two pulses where clear dips corresponding to absorption by atmospheric water vapor can be seen. \textcolor{black}{We find that the time-integrated power of the pulse emitted by PdTe$_2$ is lower than the reference pulse. However, for particular frequency bands, $f< 0.2$ THz and $1.5 < f < 2.7$ THz, the spectral intensity emitted by PdTe$_2$ is higher than the spectral intensity of the reference pulse.} This is a clear indication of nonlinear optical activity that enhances the output at these frequencies. In particular, the low-frequency, near-DC, behavior can only arise from frequency-difference processes in an even-order nonlinear response, such as optical rectification or photogalvanic or photon drag effects, since the excitation pulse does not have spectral contributions at such low frequencies\cite{Plank2018,ShenNLO}. Given the broad bandwidth of our excitation pulse, however,  multiple higher harmonic generation (HHG) orders could conceivably contribute to the observed \textcolor{black}{enhanced spectral output}, since the individual contributions cannot be clearly separated as in the visible NLO experiments.

\begin{figure}[t!]
\begin{center}
 \includegraphics[width=1\columnwidth]{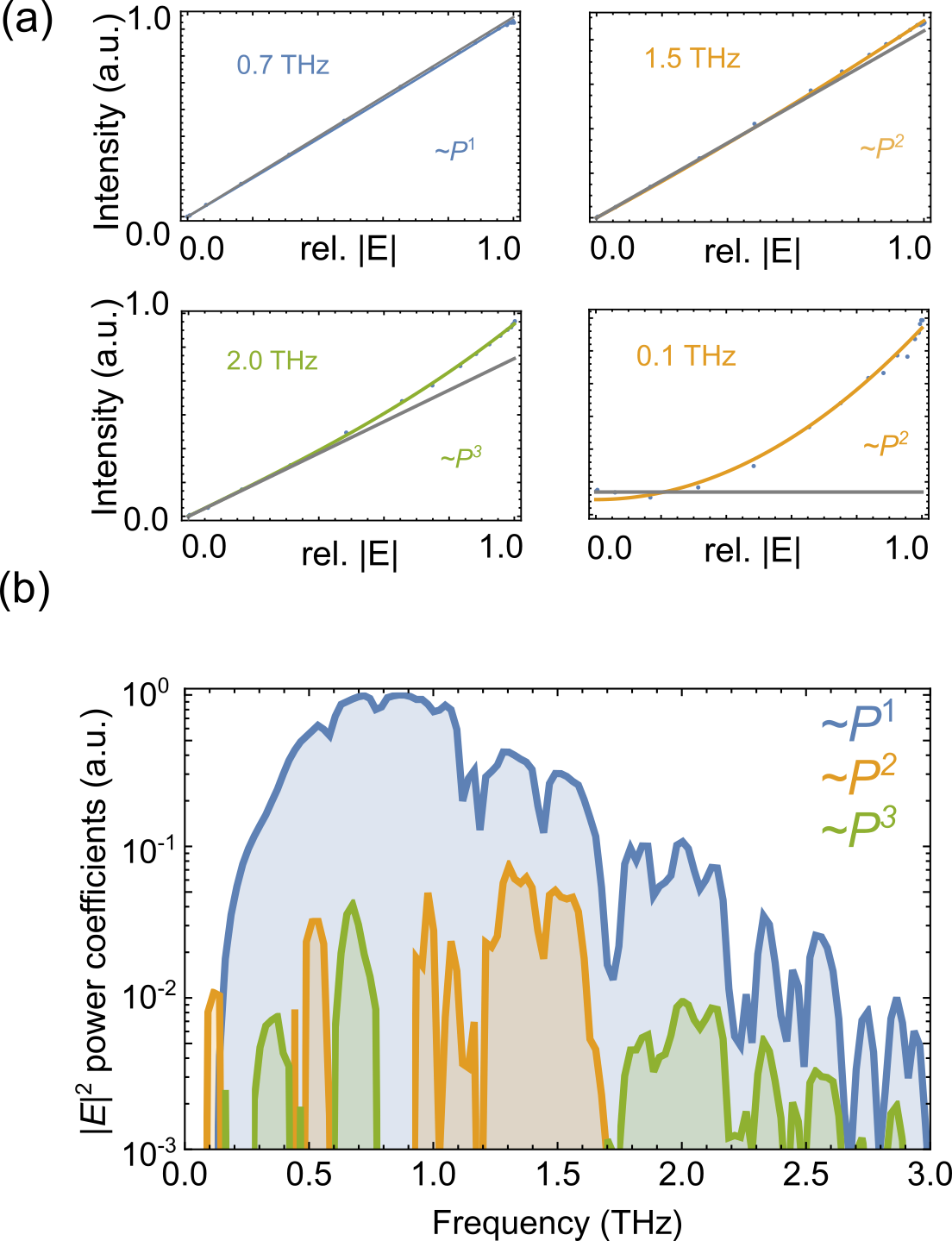}
\end{center}
 \caption{(a) Power analysis of PdTe$_2$ THz emission spectrum intensity according to Eq. \eqref{epeq} shows mixing of quadratic, linear, and cubic behavior at different frequencies. For almost all frequencies, linear dominates except towards zero frequency, where quadratic takes over. (b) A log plot of the coefficients of Eq. \eqref{epeq} shows the fitting finds either linear plus quadratic or linear plus cubic behavior at a given frequency. Linear and cubic contributions are found to dominate the 1.7 THz to 2.7 THz band where PdTe$_2$ emits greater intensities than the reference pulse, suggesting third-order processes are responsible for the observed \textcolor{black}{spectral enhancement over reference at higher frequencies}.} 
\label{fig5} 
\end{figure}

To determine which HHG orders are present within the PdTe$_2$ spectrum, we performed a power dependence study. Using a pair of polarizers placed after the THz emitter, we vary the orientation of the first while keeping the second fixed. As a result, we can fit the spectrum $E_{rad}(f)$ as a function of the power $P$ for each frequency $f$. The injected and emitted electric fields, $E_{in}(f)$ and $E_{rad}(f)$, are complex. By assuming $|E_{in}(f)| = \mathcal{E}_0 \sqrt{P}$ for some frequency independent constant $\mathcal{E}_0$ we can derive a fitting equation for the real valued spectrum as a function of power $|E_{rad}(f,P)|^2$:
\begin{eqnarray}
\begin{array}{l}
\left|E_{rad}(f,P)\right|^2 = \\
~~~~~\left|\chi^{(1)}(f) P^{1/2} + \chi^{(2)}(f) P +\chi^{(3)}(f) P^{3/2}\right|^2 ~.
\end{array} \nt\\
%     &=& \left|\chi^{(1)}(f) P^{1/2} + \chi^{(2)}(f) P +\chi^{(3)}(f) P^{3/2}\right|^2 ~.\nt\\
    \label{epeq}
\end{eqnarray}
We have now absorbed powers of $\mathcal{E}_0$ into $\chi^{(n)}$. The power dependence and corresponding fits at the pulse center frequency, as well as its multiples and in the low frequency limit, are shown in Fig. \ref{fig5}(a). The linear response, SHG, and THG coefficients $|\chi^{(n=1,2,3)}(f)|$ are shown in Fig. \ref{fig5} (b). We note that although quadratic and cubic terms are mathematically independent, over the limited available power range with dominant linear response they become nearly collinear. Consequently, the fit converges to the dominant higher-order term rather than distributing weight across both.

From this analysis, we can conclude that the \textcolor{black}{enhanced spectral output at high frequencies} in Fig. \ref{fig4}c is due to THG, and the low-frequency excess is indeed due to a second-order process. Moreover, there is a strong third-order fundamental enhancement at the THz spectral peak, indicating that for high enough powers or narrow enough THz pulses, PdTe$_2$ could be used for pulse refocusing. We note that while we can distinguish the SHG of the bare PdTe$_2$ surface from the one generated by symmetry-breaking defects and edges etc in visible experiments, the large spot size diameter of 300$\mu$m-500$\mu$m in THz experiments makes these contributions indistinguishable. However, we believe the overall SHG intensity to be dominated by the bigger area of bare PdTe$_2$ surface.

\begin{figure*}[t!]
\begin{center}
 \includegraphics[width=2\columnwidth]{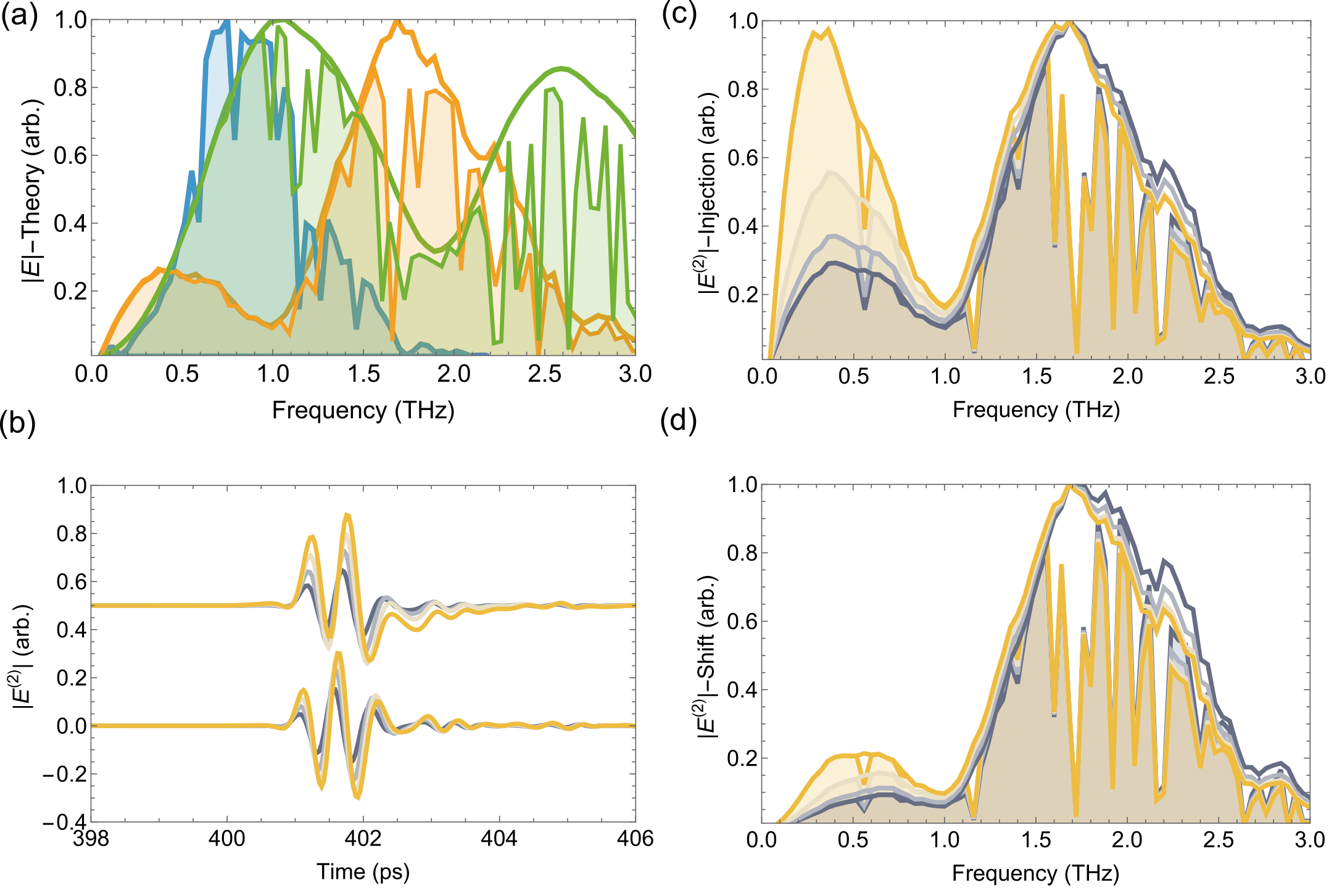}
\end{center}
 \caption{(a) Second (orange) and third (green) order response spectra calculated according to Eq.  \eqref{Eot} using the reference THz pulse (blue). Each spectrum is individually normalized by its peak intensity. The filled curves show the result of applying laboratory transmission and absorption to the theoretical spectra. (b) Time-dependent radiated pulses generated by injection (top) and shift (bottom) currents for increasing lifetimes $\tau_{inj/sh} = 0.05, 0.1, 0.2, 0.5$ ps are shown in the gray to yellow colored curves. The increasing lifetimes drive up the near DC response of the injection current. The corresponding Fourier spectra normalized by their peak intensity are shown in (c) and (d), where the filled curves account for atmospheric absorption and transmission. Phenomenologically, the injection current spectra for $\tau_{inj}$ more closely matches the data in Fig. \ref{fig5}.}
\label{fig6} 
\end{figure*}

We can compare the power analysis to the theoretically predicted HHG spectrum for the incoming pulse. The energy scales of THz spectroscopy (1 THz $\sim$ 4 meV) are far below the interband transitions we studied with visible excitations , and the technique is often employed to probe Fermi surface or intraband contributions to the Drude conductivity in quantum materials \cite{Wu2013}. At these energy and time scales, it is more natural to dynamically relate the radiated,  $E_{rad}(t)$, and incoming, $E_{in}(t)$, electric fields through the photocurrent $j(t)$ generated by the NLO conductivity. In Appendix \ref{AppB}, we show this can be written as:
\begin{eqnarray}
    j(t) &=& \left[\sigma^{(1)}*E_{in} + \sigma^{(2)}*E_{in}^2+\sigma^{(3)}*E_{in}^3\right](t)~,\nt\\
    E_{rad}(t) &\propto& - \del_t j(t;E_{in}(t))~,
\end{eqnarray}
where $*$ is the convolution operator, and $\sigma^{(n)}(t_1,\ldots,t_n)$ are the time dependent $n$th order nonlinear conductivities. 
%The dynamic content of the radiated, $E_{rad}(t)$, and incoming, $E_{in}(t)$ electric fields  are related through the photocurrent $j(t)$ via $E_{rad}(t) \propto -\del_t j(t;E_{in(t)})$. 
Following our derivation in Appendix \ref{AppB}, if the nonlinear response kernel $\sigma^{(n)}(\omega_1,\ldots,\omega_n)$ is constant over the incoming spectrum $E_{in}(\omega)$ then the $n$th nonlinear contribution to the radiated spectrum, $E^{(n)}_{rad}(t)$ is simply
\begin{eqnarray}
    E^{(n)}_{rad} (t) &\approx& - \sigma^{(n)}_0 E_{in}(t)^{n-1} 
    \cdot \del_t E_{in}(t)  ~.
    \label{Eot}
\end{eqnarray}
Here $\sigma^{(n)}_0$ is the $n$th order NLO susceptibility strength, and we have suppressed the tensor nature of the NLO response, as for linearly polarized light, all components of $E_{in}(t)$ are in phase. Using Eq. \eqref{Eot} we calculate the theoretical second and third-order emitted fields and present their Fourier spectra alongside $E_{in}(f)$ in Fig. \ref{fig5}. The theoretical spectrum indicates second and third-order activity should be dominant for the same frequencies as determined by the power analysis in Fig. \ref{fig5}. However the theory in Fig. \ref{fig6} predicts a SHG peak at $\sim$ 1.75 eV while the power analysis instead displays a peak closer to $\sim 1.5$ eV. This can be accounted for by the atmospheric (mostly water) absorption in the laboratory. The reference pulse can be seen to have a large absorption peak around 1.75 eV. To verify this, we perform direct measurement of our THz pulse in free space and then in a box with purged nitrogen atmosphere maintaining the same path length as our reflection geometry experiments to obtain the transmission spectrum of our lab \textcolor{black}{(see Appendix \ref{AppD})}. We can then apply this transmission spectrum to the theoretical higher order response, and see that the atmospherically adjusted theoretical second-order spectrum now peaks at $\sim$ 1.5 eV in  Fig. \ref{fig6}(a).

To obtain a greater understanding of the underlying excitation processes of the experimental observations, physically realistic nonlinear conductivity kernels have to be employed in the theoretical analysis. For example, in Appendix \ref{AppC} we demonstrate that by properly employing the Drude conductivity for the linear response component we see that the \textit{reflected} and incoming pulses are spectrally similar, and in fact become identical in the clean (i.e. long scattering time) limit. In the case of second-order response, we seek to elucidate why the peak values of the DC and SHG contributions to $E_{rad}$ are comparable. The true $\sigma^{(2)}(\omega_1,\omega_2)$ function will not be spectrally flat due to resonant transitions and excitation lifetimes. Consider the second-order shift and injection current response functions introduced by Braun \textit{et al}. \cite{Braun2016}:
\begin{widetext}
\begin{eqnarray}
    \sigma^{(2)}_{inj}(t-t_1,t-t_2) &=& \sigma_{inj} \delta(t_1-t_2) \Theta(t-t_1) e^{-(t-t_1)/\tau_{inj}}~, \\
\sigma^{(2)}_{sh}(t-t_1,t-t_2) &=& \sigma_{sh} \delta(t_1-t_2) \left( \delta(t-t_2) - \Theta(t-t_1) \frac{e^{-(t-t_1)/\tau_{sh}}}{\tau_{sh}}\right)~.
\end{eqnarray}
\end{widetext}
In this approximation, only the intensity of the electric field $|E(t)|^2$ matters due to the delta-function constraint $t_1 = t_2$. The relaxation times and strengths of the injection and shift excitations are given respectively by $\tau_{inj/sh}$ and $\sigma_{inj/sh}$. We apply these kernels to our injected THz pulse and take the time derivative to generate the emitted pulses in Fig. \ref{fig6}(b). We consider relaxation times of $\tau_{inj/sh} = $ 0.05, 0.1, 0.2 and 0.5 ps to cover the orders of excitation timescales observable in THz spectroscopy \cite{Braun2016}. The longer lifetimes translate into longer-lived second-order pulses, and correspondingly increase the weight of the near DC peak in the corresponding Fourier spectra in Figs. \ref{fig6}(c) and (d). In the limit of a long shift current lifetime, $\tau_{sh} \to \infty$, the shift current response kernel approaches $\sigma^{(2)}_{sh}(t-t_1,t-t_2) \to \delta(t-t_1)\delta(t-t_2) $, which is equivalent to the simple theory approach discussed previously. This explains why the shift current and simple theory spectra in Figs. \ref{fig6} (a) and (d) appear similar for long lifetimes. On the other hand, the injection current is dominated by the low frequency/DC response as the lifetime increases, suggesting that the second-order activity extracted in Fig. \ref{fig5}(b) originates from an injection current process with a lifetime on the order of 100 fs. 

\textcolor{black}{We note that while the shift and injection current kernels conceived by Braun \textit{et al.} seemingly capture the behavior of second-order spectra, at the THz energy scale excitations about the Fermi surface (including quantum geometric contributions) can play a role in the nonlinear THz response of a centrosymmetric material with topological surface states \cite{Stensberg2024,Gao2025}. Additionally, photon drag effects can manifest at the THz scale \cite{Cheng2023,Plank2016}. The microscopic formulae for these variations contributions have been determined in other works, and it has been found that in the semiclassical limit many of these responses take on the same dynamical functional behavior as formulated by Braun \textit{et al} (especially when the constraint $t_1=t_2$ is applied) \cite{Xie2025,Parker2019}. Future studies of PdTe$_2$ should thus combine microscopic Hamiltonian based calculations with transmission THz spectroscopy measurements in purged environments to isolate the different $\sigma^{(n)}_{ijk\ldots}$ tensor components, and investigate their precise origin.}

\section{Conclusion}
We have studied the NLO responses of single crystal PdTe$_2$ using visible (SHG and FWM) and THz spectroscopy. We conclude that second and third harmonic processes are clearly present in these disparate bands, indicating broadband nonlinear activity in PdTe$_2$ despite its inversion symmetric crystal structure. 

In the visible studies we found the NLO response adheres to the surface point group symmetry $C_{3v}$ of PdTe$_2$, which is is consistent with the single crystal nature of the material. Moreover, despite the inversion symmetry of the bulk crystal, clear SHG was observed and was resonantly enhanced when the SHG energy coincided with the separation of topological surface state Dirac points. Such an enhancement between buried conduction and valence band Dirac points has previously been witnessed in topological insulator photogalvanic effect studies, and suggests PdTe$_2$ hosts a circular photogalvanic effect measurable in future studies \cite{Pan2017}.

In the THz spectroscopy studies, we found that linear response combined with third harmonic generation is sufficiently strong to produce a \textcolor{black}{relative increase} in the emitted spectrum at higher frequencies. Given that third harmonic generation is the sister process of fundamental enhancement, this suggests that PdTe$_2$ could be used for self-focusing and frequency upconversion in future THz-based studies. By studying the power dependence of our radiated spectrum, we were further able to extract both second and third-order NLO processes, despite the broadband nature of the excitation pulse. Leveraging data on the transmission spectrum of our laboratory and careful application of nonlinear photocurrent theory, we found we could phenomenologically reproduce the experimental results using only the measured input pulse. To the best of our knowledge, the extraction of NLO activity in open atmosphere reflection geometry experiments is rarely conducted due to the challenges we overcame in this study. Our work lays the foundation for future THz reflection geometry experiments where transmission is not feasible, and provides a framework for studying NLO activity in broad spectra. Finally, the presence of SHG and near DC THz peaks in the emitted spectra shows that PdTe$_2$ likely possesses decent low-energy photogalvanic or photon drag effects. This motivates future studies on photocurrent generation in PdTe$_2$ to identify THz nonlinear photocurrent generation and possible nonlinear Hall effects. Particularly, the dependence of the photocurrent on the helicity of the incoming optical excitation controlled through a quarter-wave plate could clarify the injection versus shift current nature of the low-frequency second-order response. 

\begin{acknowledgments}
 The authors thank Elisabeth Schreiner and Bärbel Zimmermann for performing the ToF-SIMS measurements of the PtSe$_2$ crystal.
 GJdC would like to acknowledge funding from the US Office of the Deputy Assistant Secretary of the Army for Defense Exports and Cooperation Engineering and Scientist Exchange Fellowship program. 
 LL and AH were funded by the Deutsche Forschungsgemeinschaft (DFG, German Research Foundation) - 558231736. 
 Z.S. was supported by ERC-CZ program (project LL2101) from Ministry of Education Youth and Sports (MEYS) and by the project Advanced Functional Nanorobots (reg. No. CZ.$02.1.01/0.0/0.0/15\_003/0000444$ financed by the EFRR).
 This research was supported by a Laboratory University Collaborative Initiative award provided by the Basic Research Office in the Office of the Under Secretary of Defense for Research and Engineering and by the Army Research Office and was accomplished under Cooperative Agreement Number W911NF2520010 (STEP-TWO). The views and conclusions contained in this document are those of the authors and should not be interpreted as representing the official policies, either expressed or implied, of the Army Research Office or the U.S. Government. The U.S. Government is authorized to reproduce and distribute reprints for Government purposes notwithstanding any copyright notation herein. 
 We thank dtec.bw—Digitalization and Technology Research Center of the Bundeswehr for support (project VITAL-SENSE). dtec.bw is funded via the German Recovery and Resilience Plan by the European Union (NextGenerationEU)
 
 \end{acknowledgments}
\onecolumngrid
\newpage
\appendix

\section{SHG Area Scan and Defects}
\label{AppA}

Measurements realized in the bright spot area and the analyses of the second-harmonic signals generated by these regions in the sample of PdTe$_2$.

\begin{figure}[ht!]
\begin{center}
 \includegraphics[width=1\columnwidth]{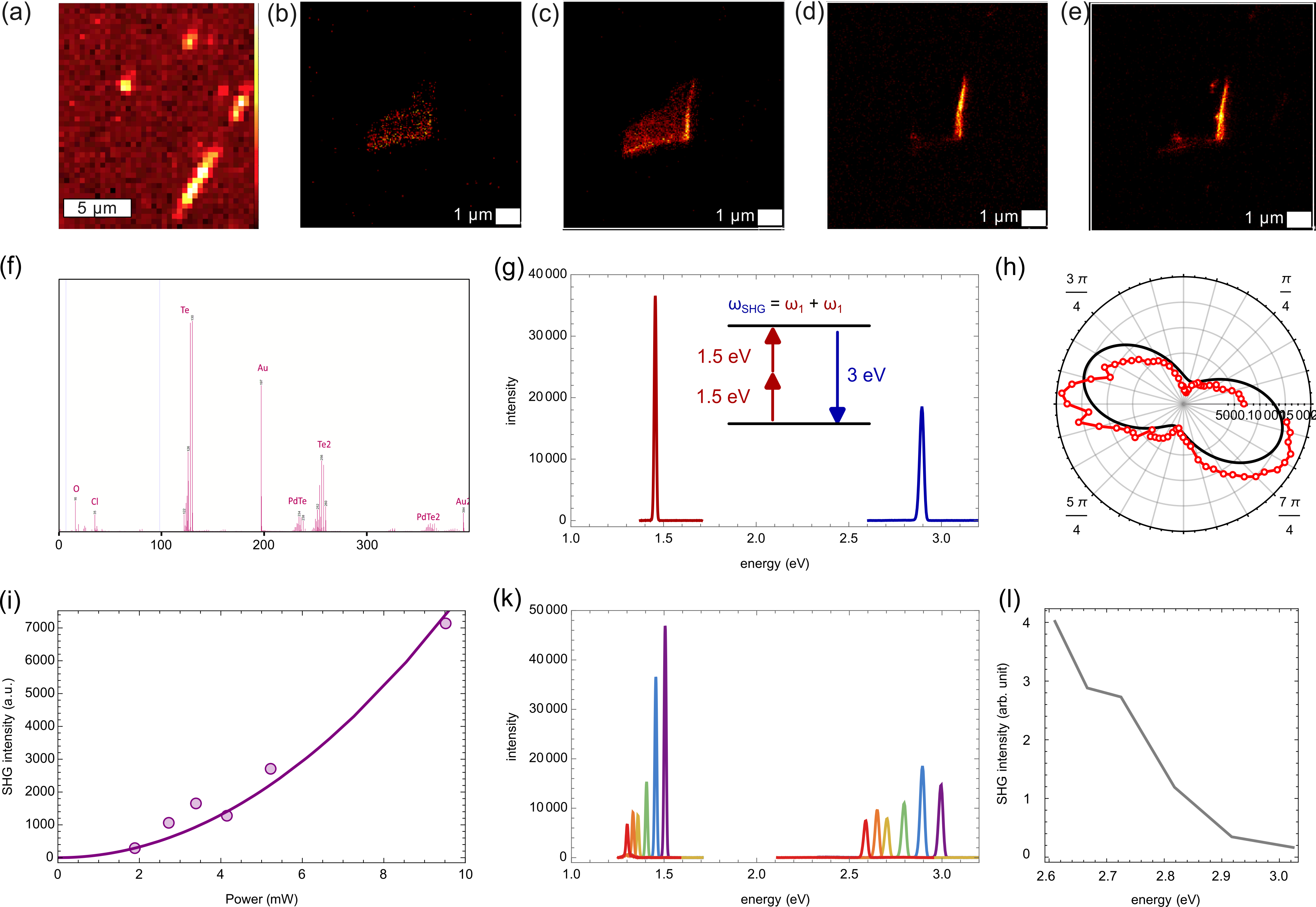}
\end{center}
 \caption{(a) Spatial scan of the SHG intensity on the PdTe$_2$ crystal surface over a 14$\times$16 $\mu$m$^2$ area. (b)-(e) Time of flight secondary ion mass spectroscopy (ToF-SIMS) intensity of detected sputtered molecular composition of an exfoliated PdTe$_2$ from the same bulk crystal showing additional clusters of chlorine and oxygen accumulation in single spots and along step edges in the PdTe$_2$. Scale bars are 1 $\mu$m  (f) Single ToF-SIMS spectrum showing detected impurities from the PdTe$_2$ crystal surface. (g) Optical second harmonic generation at 3 eV in PdTe$_2$ stimulated by a 1.5 eV pulsed laser. (h) The symmetry of the polarization of the SHG response to varying the polarization of the 1.5 eV pulse and 3 eV emission by a half-wave plate is found to exhibit two-fold symmetry which is inconsistent with the C$_{3v}$ surface point group of PdTe$_2$ but points towards a symmetry reduction due to defects and step edges at the surface.  (i) The power dependence of the SHG at 3 eV is shown to be quadratic in the pulse power, consistent with SHG scaling. The input pulse wavelength is varied in (k) and the resulting SHG intensity ratio to square of the input intensity is plotted in (i) in gray. We note that while we can observe that these brighter spots and edges can have an order of magnitude higher SHG intensity as compared to the bare PdTe$_2$ surface, we believe the overall SHG intensity to be dominated by the bigger area of bare PdTe2$_2$ surface. However, in our THz measurements, the large spot size diameter of 300$\mu$m-500$\mu$m makes these contributions indistinguishable.}
\label{App1} 
\end{figure}

\section{\textcolor{black}{Determining second- and third-order nonlinear susceptibilities ($\chi^{(2)}$ and $\chi^{(3)}$)}}
\label{Appchi}

\textcolor{black}{To determine the nonlinear optical susceptibilities of PdTe$_2$, we employed a comparative method using well–characterized reference materials measured under identical experimental conditions similar to Malard \textit{et al.} \cite{malard_observation_2013} and Lafeta \textit{et al.}\cite{Lafeta_2017}. Both the second–order and third–order responses were collected in a back-scattering geometry using the same 0.95 NA air objective, ensuring that the focal volume, collection efficiency, and Fresnel factors are directly comparable between samples. The corresponding susceptibility can then be estimated following the derivation of Boyd via\cite{Boyd}:}
\textcolor{black}{
\begin{equation}
\chi_{\mathrm{surf}}^{(2)}\!\left(R, n_S(\omega), n_S(2\omega)\right) = 2.8 \times 10^{-18} \sqrt{ \frac{0.62~R} { \left\lvert \frac{2 n_S(2\omega)}{1 + n_S(2\omega)} \right\rvert^{2} ~ \left\lvert \frac{2}{1 + n_S(\omega)} \right\rvert^{4} }} ~ \text{m}^2/\text{V}~.
\end{equation}
\textbf{Meaning of each term:}
\begin{itemize}
\item $R = \dfrac{I_{\mathrm{PdTe}_2}(2\omega)}{I_{\mathrm{Quartz}}(2\omega)}$ 
(the measured SHG intensity ratio)
\item $n_S(\omega)$: refractive index of $\mathrm{PdTe}_2$ at the fundamental wavelength \cite{Chen2022} 
\item $n_S(2\omega)$: refractive index of $\mathrm{PdTe}_2$ at the SHG wavelength \cite{Chen2022} 
\item All other constants (crystalline quartz second-order susceptibility($\chi^{(2)}_{\mathrm{CQ}} = 0.8\times 10^{-12}\,\mathrm{m}/\mathrm{V}$\cite{Lafeta_2021}), confocal interaction length, Fresnel factors for quartz) are already absorbed into the numerical prefactors $2.8 \times 10^{-18}$ and $0.62$.
\end{itemize}
}

\textcolor{black}{
The optical second-order nonlinear surface susceptibility that we report is similar to, and in some cases surpasses, the second-order nonlinear susceptibility sheet of symmetry-broken two-dimensional materials, as shown in the Table \ref{tab1}.}
\begin{table}[h!]
\centering
\caption{Second-Order Nonlinear Susceptibility of Various monolayers (ML) and heterobilayers (HBL) of 2D Materials.}
\label{tab1}
\renewcommand{\arraystretch}{1.2} 
\begin{tabular}{l c c l}
\textbf{Material} &  $\lambda_F$ \textbf{(nm)} & $\chi^{(2)}(2\omega)$ \textbf{(nm$^2$/V)} & \textbf{Ref.} \\
%& \textbf{(nm)} & \textbf{(nm$^2$/V)} & \\
\hline
ML WS$_2$  & 832 & 5.9 & \cite{Janisch_2014} \\
ML MoS$_2$ & 810 & 3.3 & \cite{Kumar_2013} \\
ML MoSe$_2$ & $\sim$800 & 1.7 & \cite{Puri2024} \\
PdTe$_2$  & 820 & 5.0 & Our work \\
HBL MoSe$_2$/WS$_2$ & 1030 & 0.0091 & \cite{Zhu2025} \\
ML WSe$_2$ & 1550 & 0.07 & \cite{Rosa2018} \\
Graphene on Au & 780 & 1.4 & \cite{Lobet2016} \\
\hline
\end{tabular}
\end{table}

\textcolor{black}{Unlike $\chi^{(2)}$, bulk inversion symmetric materials have non-zero $\chi^{(3)}$ tensor elements, and so the third-order response will be generated by multiple layers. The effective third–order susceptibility is given by the equation $\chi^{(3)}_{\mathrm{eff}} = \chi^{(3)}_{\mathrm{sheet}}/t$ and is extracted by comparing the FWM signal of PdTe$_2$ bulk with that of a 0.11 mm fused–quartz reference, whose susceptibility 
$\chi^{(3)}_{\mathrm{FQ}} = 2.0\times 10^{-22}\,\mathrm{m}^2/\mathrm{V}^2$ is 
well established \cite{Lafeta_2021}. For FWM process 
$2\omega_1 - \omega_2 = \omega_{\mathrm{FWM}}$ with 
$\lambda_1 = 780$\,nm, $\lambda_2 = 930$\,nm, and 
$\lambda_{\mathrm{FWM}} \approx 670$\,nm, the resulting expression is:
\begin{equation}
\chi^{(3)}_{\mathrm{PdTe_2}}(R,n_1,n_2,n_F)
=
2.0\times 10^{-22}\;
\frac{1.45}{n_1}\;
\sqrt{
\frac{
0.414\,R
}{
\left|\frac{2 n_F}{1+n_F}\right|^{2}
\left|\frac{2}{1+n_1}\right|^{4}
\left|\frac{2}{1+n_2}\right|^{2}
}
}
\quad \mathrm{m}^2/\mathrm{V}^2~.
\end{equation}
\textbf{Meaning of each term:}
\begin{itemize}
\item $R = I^{\mathrm{FWM}}_{\mathrm{PdTe_2}}/I^{\mathrm{FWM}}_{\mathrm{FS}}$, 
(the measured FWM intensity ratio)
\item $n_1 = n_{\mathrm{PdTe_2}}(780\,\mathrm{nm})$, 
$n_2 = n_{\mathrm{PdTe_2}}(930\,\mathrm{nm})$, and 
$n_F = n_{\mathrm{PdTe_2}}(670\,\mathrm{nm})$: refractive index of $\mathrm{PdTe}_2$ at respective wavelength \cite{Chen2022} 
\item The numerical constants incorporate the fused quartz refractive index, its 
known $\chi^{(3)}_{\mathrm{FQ}} = 2.0\times 10^{-22}\,\mathrm{m}^2/\mathrm{V}^2$, and the fixed Fresnel factors for backscattering.
\end{itemize}
Using the experimentally measured ratio $R \approx 42$ and the refractive 
indices $n_1 = 2.2$, $n_2 = 2.6$, and $n_F = 1.7$ \cite{Chen2022}, we obtain
\begin{equation}
\chi^{(3)}_{\mathrm{PdTe_2}}
\approx 2\times 10^{-21}\,\mathrm{m}^2/\mathrm{V}^2~.
\end{equation}
This value is between one and two orders of magnitude lower than third–order 
susceptibilities reported previously for transition–metal dichalcogenides such as MoTe$_2$ \cite{acs-chi-mote2}, group-VI \cite{Lafeta_2021,acs-chi-mote2, autere-chi}, and graphene \cite{Lafeta_2017,acs-chi-mote2}, confirming that PdTe$_2$ exhibits a comparable or lower third–order nonlinear optical response when compared to literature results using approximately the same wavelength (but occasionally different spectroscopy studies such as THG).
}

\section{Nonlinear Optical Response Functions and Instant-Interactions}
\label{AppB}

\begin{figure}[ht!]
\begin{center}
 \includegraphics[width=1\columnwidth]{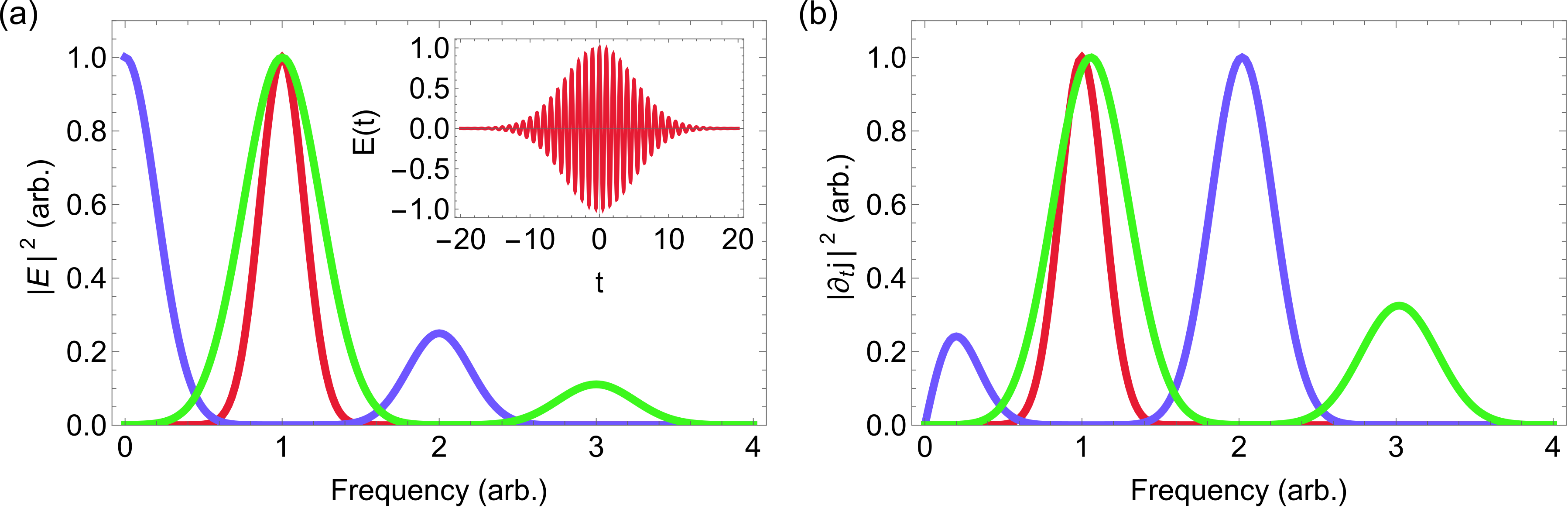}
\end{center}
 \caption{Frequency spectrum plots of $E_{rad}^{(n)}$ intensities  according to (a) the standard polarization framework $E_{rad}^{(n)} \equiv P^{(n)}(\Omega) \sim \left|  \chi_0^{(n)} \int dt e^{i\Omega t} E(t)^n \right|$ and (b) the radiative photocurrent framework in Eq. \eqref{eoutfeq}. The incoming spectrum is shown in red, and second- and third-order response spectra are shown in blue and green, respectively. The inset in both cases is incoming the monochrome pulse. The primary differences due to the extra $\Omega$ in Eq. \eqref{eoutfeq} are the elimination of the DC component, the shifted relative peak height and peak positions.}
\label{AppHHG} 
\end{figure}

The temporal representations of linear, second-order, and third-order photocurrents are given by \cite{Boyd}:
\begin{subequations}
\begin{eqnarray}
    j^{(1)}(t) &=& \int_{-\infty}^\infty E(t_1) \sigma^{(1)}(t-t_1)dt_1~, \\
    j^{(2)}(t) &=& \int_{-\infty}^\infty E(t_1)E(t_2) \sigma^{(2)}(t-t_1,t-t_2)dt_1 dt_2~,\\
    j^{(3)}(t) &=& \int_{-\infty}^\infty E(t_1)E(t_2)E(t_3)\sigma^{(3)}(t-t_1,t-t_2,t-t_3)dt_1 dt_2dt_3~.     
\end{eqnarray}
\end{subequations}
In frequency space there is always one less integral because of the convolution theorem:
\begin{subequations}
\begin{eqnarray}
    j^{(1)}(\Omega) &=&  E(\Omega) \sigma^{(1)}(\Omega)~, \\
    j^{(2)}(\Omega) &=& \int_{-\infty}^\infty E(\Omega-\omega_1)E(\omega_1) \sigma^{(2)}(\Omega-\omega_1,\omega_1)d\omega_1~,\\
    j^{(3)}(\Omega) &=& \int_{-\infty}^\infty E(\Omega-\omega_1-\omega_2)E(\omega_1)E(\omega_2)\sigma^{(3)}(\Omega-\omega_1-\omega_2,\omega_1,\omega_2)d\omega_1 d\omega_2~.
\end{eqnarray}
\end{subequations}
We have absorbed normalization constants into our definitions of $\sigma^{(n)}$ and suppressed tensor indices as for now we are only concerned with time and frequency structures. 

Within this paper, we study the electric field emitted by PdTe$_2$ after excitation with a THz pulse. If we assume the emitted electric field is generated by the acceleration and deceleration of charge carriers, then the THz field radiated from the sample and measured at location $\bm{r}$ and time $t$  can be expressed as \cite{Pettine2023}:
\begin{eqnarray}
    \bm{E}_{rad}(\bm{r},t) = - \frac{1}{4\pi \epsilon_0 c^2} \int d\bm{r}' \frac{1}{|\bm{r}-\bm{r}'|} \partial_t \bm{j}(\bm{r}',t -|\bm{r}-\bm{r}'|/c)~. \label{EradEqn}
\end{eqnarray}
Equation \eqref{EradEqn} implies that the dynamics of the measured THz field are dictated by the NLO susceptibility tensors $\chi^{(n)}(t-t_1,\dots,t-t_n)$ which are defined by:
\begin{equation}
    \chi^{(n)}(t-t_1,\dots,t-t_n) = -\frac{\del}{\del t}\sigma^{(n)}(t-t_1,\ldots,t-t_n)~.
\end{equation}
Given that 1 THz frequency light has energy of 4.14 meV, the excitations involved in THz spectroscopy are small on the scale of the PdTe$_2$ band structure shown in Fig. \ref{fig1}. This suggests that the semiclassical/low-excitation energy limits of NLO tensor theory can account for our observed phenomena. In this limit, as long as there is no resonant excitation, the Fourier representation of the NLO functions $\sigma^{(n)}(\omega_1,\ldots,\omega_n)$ will depend frequencies through terms such as $1/(\omega_n + i\gamma)$ where $\gamma = 1/\tau_s$ is the inverse lifetime of the carrier \cite{Avdoshkin2020,Onishi2024,Parker2019}. Provided $\gamma$ is large compared to the frequencies $\omega$ for which $E(\omega)$ is non-zero, we can safely approximate constant spectral behavior, $\sigma^{(n)}(\omega_1,\ldots,\omega_n) \approx const.$. In this case $\sigma^{(n)}(t-t_1,\dots,t - t_n) = \sigma_0^{(n)} \delta(t-t_1)\cdots \delta(t-t_n) $. This behavior  reflects that with respect to the timescales of the measurement, we expect the electric field to instantaneously interact with itself with some weight $\sigma_0^{(n)}$. Deviations from this base case should demonstrate a retarded response of the material owing to finite lifetimes. Then the NLO current is simply:
\begin{eqnarray}
    j^{(n)}(t) = \sigma_0^{(n)} E(t)^n~,
\end{eqnarray}
and the $n$th NLO component of the radiated field in time and Fourier space is given by
\begin{eqnarray}
    E_{rad}^{(n)}(t) &=& -n \sigma_0^{(n)}  E(t)^{n-1}\cdot \del_t E(t) ~,\\
    |E_{rad}^{(n)}(\Omega)| &=& \left|  n \sigma_0^{(n)} \int dt e^{i\Omega t} \Omega E(t)^n \right|~. \label{eoutfeq}
\end{eqnarray}

As an example, we can calculate $E_{rad}^{(2)}(\Omega)$ and $E_{rad}^{(3)}(\Omega)$ according to Eq. \eqref{eoutfeq} in response to a monochrome pulse centered at a frequency of 1. We present these in Fig. \ref{AppHHG} and compare them to the traditional polarization formalism of higher harmonic generation, where the NLO response tensor $\chi^{(n)}$ is assumed to be spectrally independent and so $E_{rad}^{(n)}(\Omega) \equiv P^{(n)}(\Omega) \sim \left|  \chi_0^{(n)} \int dt e^{i\Omega t} E(t)^n \right|$.

\begin{figure*}[b!]
\begin{center}
 \includegraphics[width=1\columnwidth]{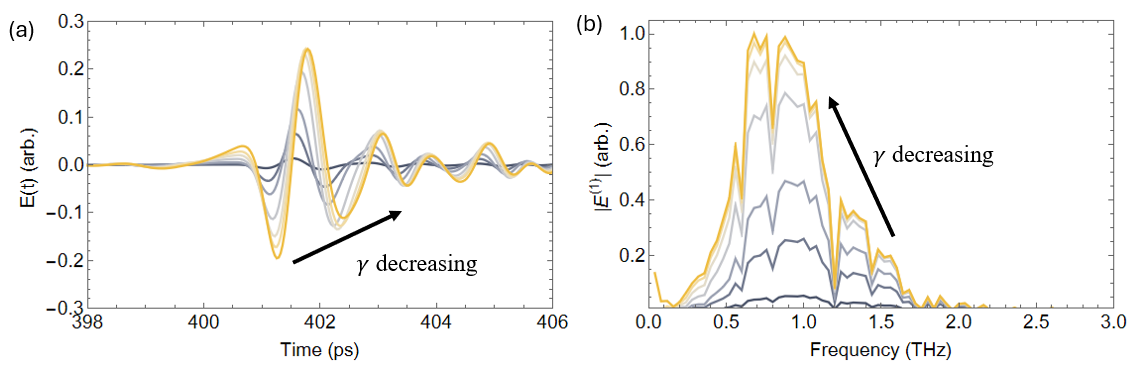}
\end{center}
 \caption{
Effects of carrier dynamics on the radiated electric field with $\gamma = 1/\tau_s$ for $\tau_s$ ranging from 0 to 1 ps in linear response. The final yellow curves correspond to the injected THz pulse used to compute the first order response according to Eq. \eqref{rado1}}
\label{App2} 
\end{figure*}

\section{Finite Lifetime In Linear Response}
\label{AppC}
It is interesting to consider the impact of lifetime on the linear response to the THz pulse. Motivated by Refs \cite{Hafez2018} and \cite{Orenstein2015}, consider the linear response current:
\begin{equation}
    j(t) = \int_{-\infty}^\infty E(t_1) \sigma(t-t_1) dt_1~.
\end{equation}
For the Drude model, we have the well known form of conductivity:
\begin{eqnarray}
\sigma(t) &=& \frac{n e^2}{m} \Theta(t) e^{-\gamma t}~, \\
\sigma(\omega) &=& \frac{n e^2}{m} \frac{- i}{\omega + i \gamma}~,
\end{eqnarray}
where $\gamma$ is the scattering rate or inverse lifetime. Multiple NLO responses such as electrical polarization, magnetization and electric quadrupole densities can underlie the current $j(t)$ further complicating the expression for $\sigma(t)$, but for now we will limit ourselves to a carrier dynamics framework \cite{annrev2021}. Following the discussion in Appendix \ref{AppB}, we relate the photocurrent to the electric field simply through  $-\partial_t j(t) \sim E_{rad}(t)$:
\begin{eqnarray}
    -\partial_t j(t) &=& -\int_{-\infty}^\infty  E(t_1) \partial_{t}\sigma(t-t_1) dt_1 ~.
\end{eqnarray}
 Then we have
\begin{equation}
    \chi(\tau) \equiv -\del_\tau \sigma(\tau) = \frac{n e^2}{m} \left( -\delta(\tau) + \gamma e^{-\gamma \tau} \Theta(\tau) \right)
\end{equation}
We have neglected the first instance of $e^{-\gamma \tau}$ as the delta function forces $\tau=0$. Temporally the radiated electric field thus has an instant response given by a first time, and a time lagged response accounting for carrier dynamics. In Ref \cite{Hafez2018} a similar expression is derived where effects such as hot electrons are accounted for through more complicated time-dependent $\gamma$. In a short lifetime approximation, $\gamma \to \infty$ and the term $\gamma e^{\gamma \tau} \to \delta(\tau)$ for $\tau\leq 0$, which ultimately eliminates the response, while for longer lifetimes $e^{-\gamma \tau} \to 0$, and the response function is just proportional to $\delta(\tau)$. This behavior is summarized in Fig. \ref{App2} where we numerically compute 
\begin{equation}
    E_{rad}(t) \sim - \int_{-\infty}^\infty E(t-t_1) \chi(t_1) dt_1 ~,
    \label{rado1}
\end{equation}
which allows us to limit the range of integration to $|t_1|<5/\gamma$ and capture 0.993 of the integral kernel. As $\tau_s \to \infty$ or $\gamma \to 0$, the reflected pulse retards and sharpens spectrally. The final yellow curves in Figs. $\ref{App2}$ (a) and (b) corresponds to the injected pulse, showing that in the clean limit the Drude model results in perfect reflection. In frequency space this conclusion could also have been reached by noting that under the Fourier transform $\del_t \sigma(t) \to i \omega \sigma(\omega)$ and when $\gamma\to 0$ one has $\omega \sigma(\omega) \to const.$, so then $E_{rad}(\omega) \propto E_{in}(\omega)$.

\section{Laboratory Absorption and Transmission}
\label{AppD}

\begin{figure}[b!]
\begin{center}
 \includegraphics[width=1\columnwidth]{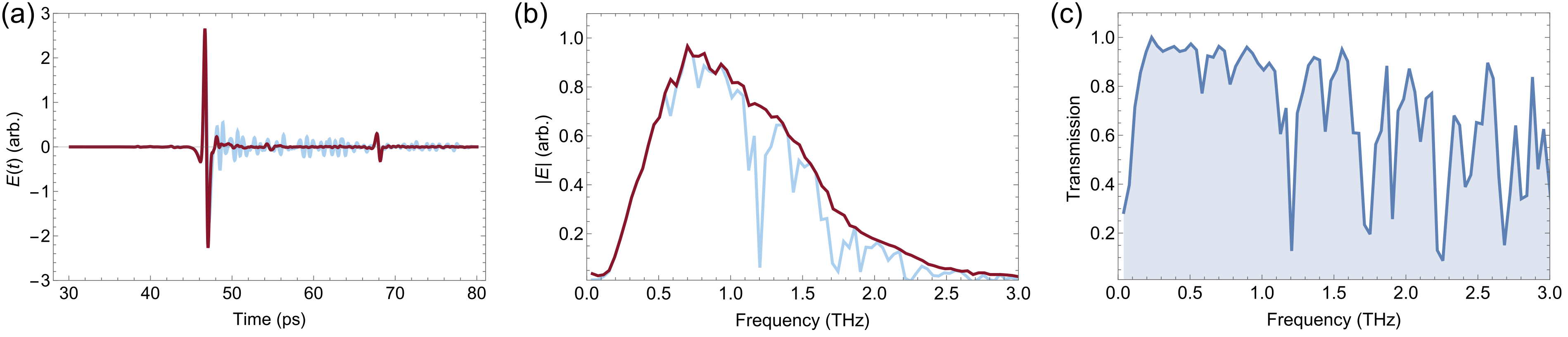}
\end{center}
 \caption{(a) THz waveform collected in purged (red) and atmospheric (blue) conditions to compute the overall transmission spectrum of the laboratory. The corresponding spectra and resulting transmission are shown in (b) and (c). }
\label{App3} 
\end{figure}

\twocolumngrid
\bibliography{references}

@article{Zhussupbekov2021,
  title = {Imaging and identification of point defects in PtTe$_2$},
  author = {Zhussupbekov, Kuanysh and Ansari, Lida and McManus, John B. and Zhussupbekova, Ainur and Shvets, Igor V. and Duesberg, Georg S. and Hurley, Paul K. and Gity, Farzan and Coile\'ain, Cormac and McEvoy, Niall},
  journal = {npj 2D Mater. Appl.},
  volume = {5},
  number = {1},
  eid = {6},
  pages = {6},
  year = {2021},
  doi = {10.1038/s41699-020-00196-8}
}

@article{McManus2020,
  title = {Low-temperature synthesis and electrocatalytic application of large-area PtTe$_2$ thin films},
  author = {McManus, John B. and Horvath, Dominik V. and Browne, Michelle P. and Cullen, Conor P. and Cunningham, Graeme and Hallam, Toby and Zhussupbekov, Kuanysh and Mullarkey, Daragh and \'O Coile\'ain, Cormac and Shvets, Igor V. and Pumera, Martin and Duesberg, Georg S. and McEvoy, Niall},
  journal = {Nanotechnology},
  volume = {31},
  number = {37},
  eid = {375601},
  pages = {375601},
  year = {2020},
  doi = {10.1088/1361-6528/ab9973}
}

@article{Yim2016,
  title = {High-Performance Hybrid Electronic Devices from Layered PtSe$_2$ Films Grown at Low Temperature},
  author = {Yim, Chanyoung and Lee, Kangho and McEvoy, Niall and O'Brien, Maria and Riazimehr, Sarah and Berner, Nina C. and Cullen, Conor P. and Kotakoski, Jani and Meyer, Jannik C. and Lemme, Max C. and Duesberg, Georg S.},
  journal = {ACS Nano},
  volume = {10},
  number = {10},
  pages = {9550--9558},
  year = {2016},
  doi = {10.1021/acsnano.6b04898}
}

@article{Yim2018,
  title = {Electrical devices from top-down structured platinum diselenide films},
  author = {Yim, Chanyoung and Passi, Vikram and Lemme, Max C. and Duesberg, Georg S. and \'O Coile\'ain, Cormac and Pallecchi, Emiliano and Fadil, Dalal and McEvoy, Niall},
  journal = {npj 2D Mater. Appl.},
  volume = {2},
  eid = {5},
  pages = {5},
  year = {2018},
  doi = {10.1038/s41699-018-0051-9}
}

@article{Wang2019,
  title = {Ultrafast Carrier Dynamics and Bandgap Renormalization in Layered PtSe$_2$},
  author = {Wang, Gaozhong and Wang, Kangpeng and McEvoy, Niall and Bai, Zhengyuan and Cullen, Conor P. and Murphy, Conor N. and McManus, John B. and Magan, John J. and Smith, Christopher M. and Duesberg, Georg S. and Kaminer, Ido and Wang, Jun and Blau, Werner J.},
  journal = {Small},
  volume = {15},
  number = {34},
  eid = {1902728},
  pages = {1902728},
  year = {2019},
  doi = {10.1002/smll.201902728}
}

@article{Wu2013,
  title = {A sudden collapse in the transport lifetime across the topological phase transition in (Bi$_{1-x}$In$_x$)$_2$Se$_3$},
  author = {Wu, Liang and Brahlek, M. and Vald\'es Aguilar, R. and Stier, A. V. and Morris, C. M. and Lubashevsky, Y. and Bilbro, L. S. and Bansal, N. and Oh, S. and Armitage, N. P.},
  journal = {Nat. Phys.},
  volume = {9},
  number = {7},
  pages = {410--414},
  year = {2013},
  doi = {10.1038/nphys2647}
}

@article{Wang2015,
  title = {Surface Recombination Limited Lifetimes of Photoexcited Carriers in Few-Layer Transition Metal Dichalcogenide MoS$_2$},
  author = {Wang, Haining and Zhang, Changjian and Rana, Farhan},
  journal = {Nano Lett.},
  volume = {15},
  number = {12},
  pages = {8204--8210},
  year = {2015},
  doi = {10.1021/acs.nanolett.5b03708}
}

@article{Hsieh2011,
  title = {Nonlinear Optical Probe of Tunable Surface Electrons on a Topological Insulator},
  author = {Hsieh, D. and McIver, J. W. and Torchinsky, D. H. and Gardner, D. R. and Lee, Y. S. and Gedik, N.},
  journal = {\prl},
  volume = {106},
  eid = {057401},
  pages = {057401},
  year = {2011},
  doi = {10.1103/PhysRevLett.106.057401}
}

@article{Mikhailov2014,
  title = {Quantum theory of third-harmonic generation in graphene},
  author = {Mikhailov, S. A.},
  journal = {\prb},
  volume = {90},
  number = {24},
  eid = {241301},
  pages = {241301},
  year = {2014},
  doi = {10.1103/PhysRevB.90.241301}
}

@article{Hafez2018,
  title = {Extremely efficient terahertz high-harmonic generation in graphene by hot Dirac fermions},
  author = {Hafez, Hassan A. and Kovalev, Sergey and Deinert, Jan-Christoph and Mics, Zolt\'an and Green, Bertram and others},
  journal = {Nature},
  volume = {561},
  pages = {507--511},
  year = {2018},
  doi = {10.1038/s41586-018-0508-1}
}

@article{Mao2024,
  title = {Enhancement of high-order harmonic generation in graphene by mid-infrared and terahertz fields},
  author = {Mao, Wenwen and Rubio, Angel and Sato, Shunsuke A.},
  journal = {\prb},
  volume = {109},
  number = {4},
  eid = {045421},
  pages = {045421},
  year = {2024},
  doi = {10.1103/PhysRevB.109.045421}
}

@article{Tielrooij2022,
  title = {Milliwatt terahertz harmonic generation from topological insulator metamaterials},
  author = {Tielrooij, Klaas-Jan and Principi, Alessandro and Reig, David Saleta and Block, Alexander and others},
  journal = {Light Sci. Appl.},
  volume = {11},
  eid = {315},
  pages = {315},
  year = {2022},
  doi = {10.1038/s41377-022-01008-y}
}

@article{Mrudul2021,
  title = {High-harmonic generation from monolayer and bilayer graphene},
  author = {Mrudul, M. S. and Dixit, Gopal},
  journal = {\prb},
  volume = {103},
  number = {9},
  eid = {094308},
  pages = {094308},
  year = {2021},
  doi = {10.1103/PhysRevB.103.094308}
}

@article{Soavi2018,
  title = {Broadband, electrically tunable third-harmonic generation in graphene},
  author = {Soavi, Giancarlo and Wang, Gang and Rostami, Habib and Purdie, David G. and others},
  journal = {Nat. Nanotechnol.},
  volume = {13},
  pages = {583--588},
  year = {2018},
  doi = {10.1038/s41565-018-0145-8}
}

@book{ShenNLO,
  title = {The Principles of Nonlinear Optics},
  author = {Shen, Y. R.},
  publisher = {John Wiley \& Sons},
  address = {New York},
  year = {1984},
  isbn = {0471889989}
}

@article{McIver2012,
  title = {Theoretical and experimental study of second harmonic generation from the surface of the topological insulator Bi$_2$Se$_3$},
  author = {McIver, J. W. and Hsieh, D. and Drapcho, S. G. and Torchinsky, D. H. and Gardner, D. R. and Lee, Y. S. and Gedik, N.},
  journal = {\prb},
  volume = {86},
  eid = {035327},
  pages = {035327},
  year = {2012},
  doi = {10.1103/PhysRevB.86.035327}
}

@article{Ge2023,
  title = {First-Principles Study of Structural and Electronic Properties of Monolayer PtX$_2$ and Janus PtXY (X, Y = S, Se, and Te) via Strain Engineering},
  author = {Ge, Xun and Zhou, Xiaohao and Sun, Deyan and Chen, Xiaoshuang},
  journal = {ACS Omega},
  volume = {8},
  number = {6},
  pages = {5715--5721},
  year = {2023},
  doi = {10.1021/acsomega.2c07271}
}

@article{Peng2020,
  title = {Strain engineering of 2D semiconductors and graphene: from strain fields to band-structure tuning and photonic applications},
  author = {Peng, Zhiwei and Chen, Xiaolin and Fan, Yulong and Srolovitz, David J. and Lei, Dangyuan},
  journal = {Light Sci. Appl.},
  volume = {9},
  eid = {190},
  pages = {190},
  year = {2020},
  doi = {10.1038/s41377-020-00421-5}
}

@article{Cai2025,
  title = {Giant room-temperature terahertz photothermoelectric response mediated by hot carriers at the metal-semimetal interfaces},
  author = {Cai, Miao and Zhang, Jinhua and Chen, Yuanbo and Hong, Liang and others},
  journal = {Sci. Adv.},
  volume = {11},
  number = {20},
  eid = {eadv0768},
  pages = {eadv0768},
  year = {2025},
  doi = {10.1126/sciadv.adv0768}
}

@article{Heiserer2025,
  title = {Impact of Strain in Free-Standing PtSe$_2$ in Scalable 2D MEMS},
  author = {Heiserer, Stefan and Galfe, Natalie and Loibl, Michael and others},
  journal = {Adv. Mater.},
  eid = {2412564},
  pages = {2412564},
  year = {2025},
  doi = {10.1002/adma.202412564}
}

@article{Sodemann2015,
  title = {Quantum Nonlinear Hall Effect Induced by Berry Curvature Dipole in Time-Reversal Invariant Materials},
  author = {Sodemann, Inti and Fu, Liang},
  journal = {\prl},
  volume = {115},
  eid = {216806},
  pages = {216806},
  year = {2015},
  doi = {10.1103/PhysRevLett.115.216806}
}

@article{Wu2022,
  title = {The field-free Josephson diode in a van der Waals heterostructure},
  author = {Wu, Heng and Wang, Yaojia and Xu, Yuanfeng and Sivakumar, Pranava K. and others},
  journal = {Nature},
  volume = {604},
  pages = {653--656},
  year = {2022},
  doi = {10.1038/s41586-022-04504-8}
}

@article{Shi2023,
  title = {Giant room-temperature nonlinearities in a monolayer Janus topological semiconductor},
  author = {Shi, Jiaojian and Xu, Haowei and Heide, Christian and others},
  journal = {Nat. Commun.},
  volume = {14},
  eid = {4953},
  pages = {4953},
  year = {2023},
  doi = {10.1038/s41467-023-40373-z}
}

@article{Connelly2024,
  title = {Emergence of threefold symmetric helical photocurrents in epitaxial low twinned Bi$_2$Se$_3$},
  author = {Connelly, Blair C. and Taylor, Patrick J. and de Coster, George J.},
  journal = {Proc. Natl. Acad. Sci. U.S.A.},
  volume = {121},
  number = {5},
  eid = {e2307425121},
  pages = {e2307425121},
  year = {2024},
  doi = {10.1073/pnas.2307425121}
}

@article{Pan2017,
  title = {Helicity dependent photocurrent in electrically gated (Bi$_{1-x}$Sb$_x$)$_2$Te$_3$ thin films},
  author = {Pan, Yu and Wang, Qing-Ze and Yeats, Andrew L. and others},
  journal = {Nat. Commun.},
  volume = {8},
  eid = {1037},
  pages = {1037},
  year = {2017},
  doi = {10.1038/s41467-017-00711-4}
}

@article{Faizanuddin2024,
  title = {Surface Second Harmonic Generation from Topological Dirac Semimetal PdTe$_2$},
  author = {Faizanuddin, Syed Mohammed and Chien, Ching-Hang and Chan, Yao-Jui and Liu, Si-Tong and Kuo, Chia-Nung and Lue, Chin Shuan and Wen, Yu-Chieh},
  journal = {arXiv preprint arXiv:2308.09053},
  year = {2024},
  doi = {10.48550/ARXIV.2308.09053}
}

@article{Zhang2021,
  title = {Terahertz detection based on nonlinear Hall effect without magnetic field},
  author = {Zhang, Yang and Fu, Liang},
  journal = {Proc. Natl. Acad. Sci. U.S.A.},
  volume = {118},
  number = {21},
  eid = {e2100736118},
  pages = {e2100736118},
  year = {2021},
  doi = {10.1073/pnas.2100736118}
}

@article{Hu2025,
  title = {Ultrabroadband nonlinear Hall rectifier using SnTe},
  author = {Hu, Fanrui and Zhao, Pengnan and Yang, Lihuan and others},
  journal = {Nat. Nanotechnol.},
  year = {2025},
  doi = {10.1038/s41565-025-01993-2}
}

@article{Plank2018,
  title = {Infrared/terahertz spectra of the photogalvanic effect in (Bi, Sb)Te based three-dimensional topological insulators},
  author = {Plank, H. and Pernul, J. and Gebert, S. and Danilov, S. N. and others},
  journal = {Phys. Rev. Mater.},
  volume = {2},
  eid = {024202},
  pages = {024202},
  year = {2018},
  doi = {10.1103/PhysRevMaterials.2.024202}
}

@article{Xie2025,
  title = {Photon-drag photovoltaic effects and quantum geometric nature},
  author = {Xie, Ying-Ming and Nagaosa, Naoto},
  journal = {Proc. Natl. Acad. Sci. U.S.A.},
  volume = {122},
  number = {9},
  eid = {e2424294122},
  pages = {e2424294122},
  year = {2025},
  doi = {10.1073/pnas.2424294122}
}

@article{Hemmat2023,
  title = {Layer-controlled nonlinear terahertz valleytronics in two-dimensional semimetal and semiconductor PtSe$_2$},
  author = {Hemmat, Minoosh and Ayari, Sabrine and Mi\v{c}ica, Martin and others},
  journal = {InfoMat},
  volume = {5},
  number = {11},
  eid = {e12468},
  pages = {e12468},
  year = {2023},
  doi = {10.1002/inf2.12468}
}

@article{Liu2020,
  title = {Two-Dimensional Materials for Energy-Efficient Spin--Orbit Torque Devices},
  author = {Liu, Yuting and Shao, Qiming},
  journal = {ACS Nano},
  volume = {14},
  number = {8},
  pages = {9389--9407},
  year = {2020},
  doi = {10.1021/acsnano.0c04403}
}

@article{acs-chi-mote2,
author = {Ha, Seongju and Kim, Hyeonkyeong and Nam, Hyunjun and Choi, Jungseok and Chae, Kwanbyung and Lee, Jae-Ung and Park, Ji-Yong and Yoo, Youngdong and Yeom, Dong-Il},
title = {Enhanced Optical Third-Harmonic Generation in Phase-Engineered MoTe2 Thin Films},
journal = {ACS Photonics},
volume = {9},
number = {8},
pages = {2600-2606},
year = {2022},
doi = {10.1021/acsphotonics.2c00222},

URL = { 
    
        https://doi.org/10.1021/acsphotonics.2c00222
    
    

},
eprint = { 
    
        https://doi.org/10.1021/acsphotonics.2c00222
    
    

}

}

@article{autere-chi,
  title = {Optical harmonic generation in monolayer group-VI transition metal dichalcogenides},
  author = {Autere, Anton and Jussila, Henri and Marini, Andrea and Saavedra, J. R. M. and Dai, Yunyun and S\"ayn\"atjoki, Antti and Karvonen, Lasse and Yang, He and Amirsolaimani, Babak and Norwood, Robert A. and Peyghambarian, Nasser and Lipsanen, Harri and Kieu, Khanh and de Abajo, F. Javier Garc\'{\i}a and Sun, Zhipei},
  journal = {Phys. Rev. B},
  volume = {98},
  issue = {11},
  pages = {115426},
  numpages = {7},
  year = {2018},
  month = {Sep},
  publisher = {American Physical Society},
  doi = {10.1103/PhysRevB.98.115426},
  url = {https://link.aps.org/doi/10.1103/PhysRevB.98.115426}
}

@article{Hidding2023,
  title = {Role of self-torques in transition metal dichalcogenide/ferromagnet bilayers},
  author = {Hidding, Jan and M\"ertiri, Klaiv and Mujid, Fauzia and Liang, Ce and Park, Jiwoong and Guimar\~aes, Marcos H. D.},
  journal = {\prb},
  volume = {108},
  eid = {064419},
  pages = {064419},
  year = {2023},
  doi = {10.1103/PhysRevB.108.064419}
}

@article{Parker2019,
  title = {Diagrammatic approach to nonlinear optical response with application to Weyl semimetals},
  author = {Parker, Daniel E. and Morimoto, Takahiro and Orenstein, Joseph and Moore, Joel E.},
  journal = {\prb},
  volume = {99},
  eid = {045121},
  pages = {045121},
  year = {2019},
  doi = {10.1103/PhysRevB.99.045121}
}

@article{Sivakumar2024,
  title = {Long-range phase coherence and tunable second order $\phi$0-Josephson effect in a Dirac semimetal 1T-PtTe$_2$},
  author = {Sivakumar, Pranava K. and Ahari, Mostafa T. and Kim, Jae-Keun and others},
  journal = {Commun. Phys.},
  volume = {7},
  eid = {312},
  pages = {312},
  year = {2024},
  doi = {10.1038/s42005-024-01825-0}
}

@article{Avdoshkin2020,
  title = {Interactions Remove the Quantization of the Chiral Photocurrent at Weyl Points},
  author = {Avdoshkin, Alexander and Kozii, Vladyslav and Moore, Joel E.},
  journal = {\prl},
  volume = {124},
  eid = {196603},
  pages = {196603},
  year = {2020},
  doi = {10.1103/PhysRevLett.124.196603}
}

@article{Onishi2024,
  title = {High-efficiency energy harvesting based on a nonlinear Hall rectifier},
  author = {Onishi, Yugo and Fu, Liang},
  journal = {\prb},
  volume = {110},
  eid = {075122},
  pages = {075122},
  year = {2024},
  doi = {10.1103/PhysRevB.110.075122}
}

@article{Orenstein2015,
  title = {Terahertz time-domain spectroscopy of transient metallic and superconducting states},
  author = {Orenstein, J. and Dodge, J. S.},
  journal = {\prb},
  volume = {92},
  eid = {134507},
  pages = {134507},
  year = {2015},
  doi = {10.1103/PhysRevB.92.134507}
}

@article{Braun2016,
  title = {Ultrafast photocurrents at the surface of the three-dimensional topological insulator Bi$_2$Se$_3$},
  author = {Braun, Lukas and Mussler, Gregor and Hruban, Andrzej and others},
  journal = {Nat. Commun.},
  volume = {7},
  eid = {13259},
  pages = {13259},
  year = {2016},
  doi = {10.1038/ncomms13259}
}

@article{Pettine2023,
  title = {Ultrafast terahertz emission from emerging symmetry-broken materials},
  author = {Pettine, Jacob and Padmanabhan, Prashant and Sirica, Nicholas and Prasankumar, Rohit P. and Taylor, Antoinette J. and Chen, Hou-Tong},
  journal = {Light Sci. Appl.},
  volume = {12},
  eid = {150},
  pages = {150},
  year = {2023},
  doi = {10.1038/s41377-023-01163-w}
}

@article{annrev2021,
  title = {Topology and Symmetry of Quantum Materials via Nonlinear Optical Responses},
  author = {Orenstein, J. and Moore, J. E. and Morimoto, T. and Torchinsky, D. H. and Harter, J. W. and Hsieh, D.},
  journal = {Annu. Rev. Condens. Matter Phys.},
  volume = {12},
  pages = {247--272},
  year = {2021},
  doi = {10.1146/annurev-conmatphys-031218-013712}
}

@article{zheng_2018,
  title = {Detailed study of the Fermi surfaces of the type-II Dirac semimetallic candidates $X{\mathrm{Te}}_{2}$ (X=Pd, Pt)},
  author = {Zheng, W. and Sch\"onemann, R. and Aryal, N. and Zhou, Q. and others},
  journal = {\prb},
  volume = {97},
  eid = {235154},
  pages = {235154},
  year = {2018},
  doi = {10.1103/PhysRevB.97.235154}
}

@article{clark_2018,
  title = {Fermiology and Superconductivity of Topological Surface States in ${\mathrm{PdTe}}_{2}$},
  author = {Clark, O. J. and Neat, M. J. and Okawa, K. and Bawden, L. and others},
  journal = {\prl},
  volume = {120},
  eid = {156401},
  pages = {156401},
  year = {2018},
  doi = {10.1103/PhysRevLett.120.156401}
}

@article{Chu_2024,
  title = {Intense second-harmonic generation in two-dimensional PtSe$_2$},
  author = {Chu, Lingrui and Li, Ziqi and Zhu, Han and Lv, Hengyue and Chen, Feng},
  journal = {Nanophotonics},
  volume = {13},
  number = {18},
  pages = {3457--3464},
  year = {2024},
  doi = {10.1515/nanoph-2024-0107}
}

@article{Guo_2020,
  title = {Anisotropic ultrasensitive PdTe$_2$-based phototransistor for room-temperature long-wavelength detection},
  author = {Guo, Cheng and Hu, Yibin and Chen, Gang and Wei, Dacheng and others},
  journal = {Sci. Adv.},
  volume = {6},
  number = {36},
  eid = {eabb6500},
  pages = {eabb6500},
  year = {2020},
  doi = {10.1126/sciadv.abb6500}
}

@article{Yu_2021,
  title = {Giant nonlinear optical activity in two-dimensional palladium diselenide},
  author = {Yu, Juan and Kuang, Xiaofei and Li, Junzi and Zhong, Jiahong and others},
  journal = {Nat. Commun.},
  volume = {12},
  eid = {1080},
  pages = {1080},
  year = {2021},
  doi = {10.1038/s41467-021-21267-4}
}

@article{Lafeta_2021,
  title = {Second- and third-order optical susceptibilities across excitons states in 2D monolayer transition metal dichalcogenides},
  author = {Lafeta, Lucas and Corradi, Aurea and Zhang, Tianyi and others},
  journal = {2D Mater.},
  volume = {8},
  number = {3},
  eid = {035010},
  pages = {035010},
  year = {2021},
  doi = {10.1088/2053-1583/abeed4}
}

@article{Lange_2024,
  title = {Ultrafast Phase-Control of the Nonlinear Optical Response of 2D Semiconductors},
  author = {Lange, Lucas and Wang, Kunliang and Bange, Sebastian and others},
  journal = {ACS Photonics},
  volume = {11},
  number = {8},
  pages = {3112--3122},
  year = {2024},
  doi = {10.1021/acsphotonics.4c00388}
}

@Book{Boyd,
  author = {Boyd, Robert W.},
  title = {Nonlinear Optics (Third Edition)},
  publisher = {Academic Press},
  address = {Burlington},
  year = {2008},
  doi = {10.1016/B978-0-12-369470-6.00001-0}
}

@article{AliSciRep2018,
  title={Galvanomagnetic properties of the putative type-II Dirac semimetal PtTe$_2$},
  author={Ali, Mazhar N. and Schoop, Leslie M. and others},
  journal={Sci. Rep.},
  volume={8},
  eid={15383},
  pages={15383},
  year={2018},
  doi={10.1038/s41598-018-29545-w}
}

@article{SereniSciRep2016,
  title={De Haas-van Alphen and magnetoresistance reveal predominantly single-band behavior in PdTe$_2$},
  author={Das, Prabhat and Chopdekar, Rajesh V. and others},
  journal={Sci. Rep.},
  volume={6},
  eid={31554},
  pages={31554},
  year={2016},
  doi={10.1038/srep31554}
}

@article{BalicasArXiv2018,
  title={Detailed study on the Fermi surfaces of the type-II Dirac semimetallic candidates PdTe$_2$ and PtTe$_2$},
  author={Zheng, W. and Sch\"onemann, R. and Aryal, N. and Zhou, Q. and others},
  journal={arXiv preprint arXiv:1805.00087},
  year={2018},
  url={https://arxiv.org/abs/1805.00087}
}

@article{PtSe2PRL2020,
  title={Dimensionality-Mediated Semimetal--Semiconductor Transition in Ultrathin PtSe$_2$ Films},
  author={Zhao, L. and Wang, Y. and others},
  journal = {\prl},
  volume={124},
  number={3},
  eid={036402},
  pages={036402},
  year={2020},
  doi={10.1103/PhysRevLett.124.036402}
}

@article{KrasheninnikovACSNano2021,
  title={Layer-Dependent Band Gaps of Platinum Dichalcogenides},
  author={Gao, Q. and Komsa, Hannu-Pekka and Krasheninnikov, Arkady V. and others},
  journal={ACS Nano},
  volume={15},
  number={8},
  pages={12968--12975},
  year={2021},
  doi={10.1021/acsnano.1c02971}
}

@article{PtTe2Crossover2019,
  title={Crossover from 2D metal to 3D Dirac semimetal in metallic PtTe$_2$ films},
  author={Huang, Z. and Zhang, K. and others},
  journal={Mater. Today Phys.},
  volume={11},
  eid={100161},
  pages={100161},
  year={2019},
  doi={10.1016/j.mtphys.2019.100161}
}

@article{lafeta_probing_2025,
  title = {Probing Noncentrosymmetric 2D Materials by Fourier Space Second Harmonic Imaging},
  author = {Lafeta, Lucas and Hartmann, Sean and Rosa, B\'arbara and Reitzenstein, Stephan and Malard, Leandro M. and Hartschuh, Achim},
  journal = {ACS Photonics},
  volume = {12},
  number = {1},
  pages = {357--363},
  year = {2025},
  doi = {10.1021/acsphotonics.4c01724}
}

@article{li_probing_2013,
  title = {Probing Symmetry Properties of Few-Layer MoS$_2$ and h-BN by Optical Second-Harmonic Generation},
  author = {Li, Yilei and Rao, Yi and Mak, Kin Fai and You, Yumeng and Wang, Shuyuan and Dean, Cory R. and Heinz, Tony F.},
  journal = {Nano Lett.},
  volume = {13},
  number = {7},
  pages = {3329--3333},
  year = {2013},
  doi = {10.1021/nl401561r}
}

@article{autere_nonlinear_2018,
  title = {Nonlinear Optics with 2D Layered Materials},
  author = {Autere, Anton and Jussila, Henri and Dai, Yunyun and Wang, Yadong and Lipsanen, Harri and Sun, Zhipei},
  journal = {Adv. Mater.},
  volume = {30},
  number = {24},
  eid = {1705963},
  pages = {1705963},
  year = {2018},
  doi = {10.1002/adma.201705963}
}

@article{malard_observation_2013,
  title = {Observation of intense second harmonic generation from MoS$_2$ atomic crystals},
  author = {Malard, Leandro M. and Alencar, Thonimar V. and Barboza, Ana Paula M. and Mak, Kin Fai and de Paula, Ana M.},
  journal = {\prb},
  volume = {87},
  number = {20},
  eid = {201401},
  pages = {201401},
  year = {2013},
  doi = {10.1103/PhysRevB.87.201401}
}

@article{Lafeta_2017,
  title = {Anomalous Nonlinear Optical Response of Graphene Near Phonon Resonances},
  author = {Lafeta, Lucas and Cadore, Alisson R. and Mendes-de-Sa, Thiago G. and others},
  journal = {Nano Lett.},
  volume = {17},
  number = {6},
  pages = {3447--3451},
  year = {2017},
  doi = {10.1021/acs.nanolett.7b00329}
}

@article{gordeev_2023,
  title = {Excitonic Resonances in Coherent Anti-Stokes Raman Scattering from Single-Walled Carbon Nanotubes},
  author = {Gordeev, Georgy and Lafeta, Lucas and Flavel, Benjamin S. and Jorio, Ado and Malard, Leandro M.},
  journal = {J. Phys. Chem. C},
  volume = {127},
  number = {41},
  pages = {20438--20444},
  year = {2023},
  doi = {10.1021/acs.jpcc.3c05696}
}

@article{Virga_2019,
  title={Coherent anti-Stokes Raman spectroscopy of single and multi-layer graphene},
  author={Virga, A. and Ferrante, C. and Batignani, G. and De Fazio, D. and others},
  journal={Nat. Commun.},
  volume={10},
  eid={3658},
  pages={3658},
  year={2019},
  doi={10.1038/s41467-019-11165-1}
}

@article{wen2019,
  title = {Nonlinear optics of two-dimensional transition metal dichalcogenides},
  author = {Wen, Xinglin and Gong, Zibo and Li, Dehui},
  journal = {InfoMat},
  volume = {1},
  number = {3},
  pages = {317--337},
  year = {2019},
  doi = {10.1002/inf2.12024}
}

@article{Plank2016,
  title = {Photon drag effect in ${(\mathrm{Bi}_{1-x}\mathrm{Sb}_{x})}_{2}\mathrm{Te}_{3}$ three-dimensional topological insulators},
  author = {Plank, H. and Golub, L. E. and Bauer, S. and Bel'kov, V. V. and Herrmann, T. and Olbrich, P. and Eschbach, M. and Plucinski, L. and Schneider, C. M. and Kampmeier, J. and Lanius, M. and Mussler, G. and Gr\"utzmacher, D. and Ganichev, S. D.},
  journal = {Phys. Rev. B},
  volume = {93},
  pages = {125434},
  year = {2016},
  month = {Mar},
  doi = {10.1103/PhysRevB.93.125434}
}

@article{Cheng2023,
  title = {Giant photon momentum locked {THz} emission in a centrosymmetric {D}irac semimetal},
  author = {Cheng, Liang and Xiong, Ying and Kang, Lixing and Gao, Yu and Chang, Qing and Chen, Mengji and Qi, Jingbo and Yang, Hyunsoo and Liu, Zheng and Song, Justin C. W. and Chia, Elbert E. M.},
  journal = {Sci. Adv.},
  volume = {9},
  number = {1},
  pages = {eadd7856},
  year = {2023},
  doi = {10.1126/sciadv.add7856}
}

@article{Stensberg2024,
  title   = {Observation of terahertz second harmonic generation from {D}irac surface states in the topological insulator {$\mathrm{Bi}_{2}\mathrm{Se}_{3}$}},
  author  = {Stensberg, Jonathan and Han, Xingyue and Ni, Zhuoliang and Yao, Xiong and Yuan, Xiaoyu and Mallick, Debarghya and Gandhi, Akshat and Oh, Seongshik and Wu, Liang},
  journal = {Phys. Rev. B},
  volume  = {109},
  issue   = {24},
  pages   = {245112},
  year    = {2024},
  month   = {Jun},
  doi     = {10.1103/PhysRevB.109.245112}
}

@article{Gao2025,
  title   = {Strong terahertz second-harmonic generation from surface states in the topological insulator {${\mathrm{Bi}}_{2}{\mathrm{Te}}_{3}$}},
  author  = {Gao, Yu and Kang, Lixing and Cheng, Liang and Xiong, Ying and Chang, Qing and Chen, Mengji and Qi, Jingbo and Yang, Hyunsoo and Liu, Zheng and Song, Justin C. W. and Chia, Elbert E. M.},
  journal = {Nat. Nanotechnol.},
  year    = {2025},
  month   = {Feb},
  doi     = {10.1038/s41565-025-01993-2}
}

@article{Lobet2016,
  title   = {Probing Graphene ${\chi}^{(2)}$ Using a Gold Photon Sieve},
  author  = {Lobet, M. and Sarrazin, M. and Cecchet, F. and Reckinger, N. and Vlad, A. and Colomer, J.-F. and Lis, D.},
  journal = {Nano Lett.},
  volume  = {16},
  number  = {1},
  pages   = {48--54},
  year    = {2016},
  doi     = {10.1021/acs.nanolett.5b02494}
}

@article{Rosa2018,
  title   = {Characterization of the second- and third-harmonic optical susceptibilities of atomically thin tungsten diselenide},
  author  = {Rosa, H. G. and Wei, H. Y. and Verzhbitskiy, I. and Rodrigues, M. J. F. L. and Taniguchi, T. and Watanabe, K. and Eda, G. and Pereira, V. M. and Gomes, J. C. V.},
  journal = {Sci. Rep.},
  volume  = {8},
  pages   = {10035},
  year    = {2018},
  doi     = {10.1038/s41598-018-28374-1}
}

@article{Faizanuddin_arxiv_2024,
    title   = {Surface Second Harmonic Generation from Topological Dirac Semimetal {PdTe$_{2}$}},
    author  = {Faizanuddin, Syed Mohammed and Chien, Ching-Hang and Chan, Yao-Jui and Liu, Si-Tong and Kuo, Chia-Nung and Lue, Chin Shuan and Wen, Yu-Chieh},
    year    = {2023},
    journal = {arXiv preprint arXiv:2308.09053},
    eprint  = {2308.09053},
    archivePrefix = {arXiv},
    primaryClass = {cond-mat.mtrl-sci}
}

@article{Puri2024,
  title = {Substrate Interference and Strain in the Second-Harmonic Generation from {MoSe2} Monolayers},
  author = {Puri, Sudeep and Patel, Sneha and Cabellos, Jose Luis and Rosas-Hernandez, Luis Enrique and Reynolds, Katlin and Churchill, Hugh O. H. and Barraza-Lopez, Salvador and Mendoza, Bernardo S. and Nakamura, Hiroyuki},
  journal = {Nano Lett.},
  volume = {24},
  number = {41},
  pages = {13061--13067},
  year = {2024},
  doi = {10.1021/acs.nanolett.4c03880},
  publisher = {American Chemical Society}
}

@article{Zhu2025,
  title = {Determining the complex second-order optical susceptibility in macroscale van der {Waals} heterobilayers},
  author = {Zhu, Zeyuan and Yoo, Taejun and Shaikh, Kanchan and Johnson, Amalya C. and Li, Qiuyang and Liu, Fang and Deng, Hui and Kobayashi, Yuki},
  journal = {J. Chem. Phys.},
  volume = {163},
  number = {17},
  pages = {174707},
  year = {2025},
  doi = {10.1063/5.0292283}
}

@article{Chen2022,
  title   = {A quasi-2D perovskite antireflection coating to boost the performance of multilayered {PdTe}$_{2}$/{Ge} heterostructure-based near-infrared photodetectors},
  author  = {Chen, Huahan and Xie, Chao and Zhong, Xianpeng and Liang, Yi and Yang, Wenhua and Wu, Chunyan and Luo, Linbao},
  journal = {J. Mater. Chem. C},
  volume  = {10},
  number  = {15},
  pages   = {6025--6035},
  year    = {2022},
  doi     = {10.1039/D2TC00438K}
}

@article{Janisch_2014,
	doi = {10.1038/srep05530},
	url = {https://doi.org/10.1038%2Fsrep05530},
	year = 2014,
	month = {jul},
	publisher = {Springer Science and Business Media LLC},
	volume = {4},
	number = {1},
	pages = {5530},
	author = {Corey Janisch and Yuanxi Wang and Ding Ma and Nikhil Mehta and Ana Laura El{\'{\i}}as and N{\'{e}}stor Perea-L{\'{o}}pez and Mauricio Terrones and Vincent Crespi and Zhiwen Liu},
	title = {Extraordinary Second Harmonic Generation in Tungsten Disulfide Monolayers},
	journal = {Scientific Reports}
}

@article{Kumar_2013,
    doi = {10.1103/physrevb.87.161403},
    url = {https://doi.org/10.1103%2Fphysrevb.87.161403},
    year = 2013,
    month = {apr},
    publisher = {American Physical Society ({APS})},
    volume = {87},
    number = {16},
    author = {Nardeep Kumar and Sina Najmaei and Qiannan Cui and Frank Ceballos and Pulickel M. Ajayan and Jun Lou and Hui Zhao},
    title = {Second harmonic microscopy of monolayer MoS2},
    journal = {Physical Review B}
}

\end{document}